\documentclass[manuscript]{acmart}

\usepackage{subcaption}
\usepackage{float}
\usepackage{makecell}
\usepackage{mathtools}
\usepackage{multirow}
\usepackage{lscape}
\usepackage{enumitem}

\definecolor{maroon}{cmyk}{0,0.87,0.68,0.32}

\newcommand{\anonymize}[1]{MOODS}

\newcommand{\edit}[1]{#1}

\newcommand*{\SuperScriptSameStyle}[1]{%
  \ensuremath{%
    \mathchoice
      {{}^{\displaystyle #1}}%
      {{}^{\textstyle #1}}%
      {{}^{\scriptstyle #1}}%
      {{}^{\scriptscriptstyle #1}}%
  }%
}

\newcommand*{\oneS}{\SuperScriptSameStyle{*}}

\newlist{questions}{enumerate}{2}

\setlist[questions,1]{label=\bf{RQ\arabic*}.,ref=RQ\arabic*}
\AtBeginDocument{%
  \providecommand\BibTeX{{%
    \normalfont B\kern-0.5em{\scshape i\kern-0.25em b}\kern-0.8em\TeX}}}

\copyrightyear{2024}
\acmYear{2024}
\setcopyright{acmlicensed}\acmConference[CHI '24]{Proceedings of the CHI Conference on Human Factors in Computing Systems}{May 11--16, 2024}{Honolulu, HI, USA}
\acmBooktitle{Proceedings of the CHI Conference on Human Factors in Computing Systems (CHI '24), May 11--16, 2024, Honolulu, HI, USA}
\acmDOI{10.1145/3613904.3642662}
\acmISBN{979-8-4007-0330-0/24/05}

\author{Sameer Neupane}

\authornote{Corresponding Author}
\email{sameer.neupane@memphis.edu}
\affiliation{%
  \institution{University of Memphis}
  \streetaddress{}
  \city{Memphis}
  \state{Tennessee}
  \country{USA}
  \postcode{}
}

\author{Mithun Saha}

\email{msaha1@memphis.edu}
\affiliation{%
  \institution{University of Memphis}
  \streetaddress{}
  \city{Memphis}
  \state{Tennessee}
  \country{USA}
  \postcode{}
}

\author{Nasir Ali}

\email{nasir.ali08@gmail.com}
\authornote{Work done while working at the University of Memphis}

\affiliation{%
  \institution{University of Memphis}
  \streetaddress{}
  \city{Memphis}
  \state{Tennessee}
  \country{USA}
  \postcode{}
}
\author{Timothy Hnat}
\authornotemark[3]
\authornote{Timothy Hnat and Santosh Kumar have financial interests in CuesHub, PBC, including stock ownership and management roles.}
\email{tim@cueshub.com}
\affiliation{%
  \institution{CuesHub, PBC}
  \streetaddress{}
  \city{Memphis}
  \state{Tennessee}
  \country{USA}
  \postcode{}
}
\author{Shahin Alan Samiei}

\email{ssamiei@memphis.edu}
\affiliation{%
  \institution{University of Memphis}
  \streetaddress{}
  \city{Memphis}
  \state{Tennessee}
  \country{USA}
  \postcode{}
}
\author{Anandatirtha Nandugudi}
\authornotemark[3]
\email{anandatirtha@gmail.com}

\affiliation{%
  \institution{University of Memphis}
  \streetaddress{}
  \city{Memphis}
  \state{Tennessee}
  \country{USA}
  \postcode{}
}

\author{David M. Almeida}

\email{dalmeida@psu.edu}
\affiliation{%
  \institution{The Pennsylvania State University}
  \streetaddress{}
  \city{University Park}
  \state{Pennsylvania}
  \country{USA}
  \postcode{}
}
\author{Santosh Kumar}
\authornotemark[4]
\email{santosh.kumar@memphis.edu}
\affiliation{%
  \institution{University of Memphis}
  \streetaddress{}
  \city{Memphis}
  \state{Tennessee}
  \country{USA}
  \postcode{}
}

\renewcommand{\shortauthors}{Sameer, et al.}

\begin{document}

\title[Momentary Stressor Logging and Reflective Visualizations: Implications for Stress Management]{Momentary Stressor Logging and Reflective Visualizations: Implications for Stress Management with Wearables}
\titlenote{A version of this work appears in ACM CHI 2024.}



\begin{abstract}

Commercial wearables from Fitbit, Garmin, and Whoop have recently introduced real-time notifications based on detecting changes in \edit{physiological} responses indicating potential stress. 
In this paper, we investigate how these new capabilities can be leveraged to improve stress management. We developed a \edit{smartwatch} app, a smartphone app, and a cloud service, and conducted a 100-day field study with 122 participants who received prompts triggered by \edit{physiological} responses several times a day. They were asked whether they were stressed, and if so, to log the most likely stressor. Each week, participants received new visualizations of their data to self-reflect on patterns and trends. Participants reported better awareness of their stressors, and self-initiating \edit{fourteen kinds of behavioral changes} to reduce stress in their daily lives. Repeated self-reports over 14 weeks showed reductions in both stress intensity (in 26,521 momentary ratings) and stress frequency (in 1,057 weekly surveys). 

\end{abstract}

%
%






%
\maketitle

\section{Introduction}

In modern society, repeated or excessive exposure to stress is prevalent and it can have detrimental effects on physical, mental, social, and financial well-being~\cite{sapolsky2004zebras}. \edit{Stress disrupts} the body's homeostasis, triggering an overproduction of stress hormones such as cortisol and adrenaline~\cite{chrousos2009stress}. This physiological imbalance elevates the risks of chronic diseases such as diabetes, hypertension, heart disease, and cancer~\cite{dai2020chronic, chrousos2009stress}. Stress also exerts a profound influence on mental well-being, elevating anxiety, depression, poor sleep, and fatigue~\cite{guilliams2010chronic}. 
The adverse cycle perpetuated by stress can drive individuals toward unhealthy coping mechanisms like unhealthy eating, substance use, or social isolation, corroding their overall quality of life~\cite{leonard2015multi}. The ramifications of stress extend to tangible consequences such as workplace incidents, absenteeism, and reduced efficiency~\cite{TheAmericanInstituteofStress_2023}. Stress is estimated to cost over \$400 billion in the United States~\cite{hassard2014calculating} alone.

After two decades of research by the scientific community to detect stress in real-time using wearables~\cite{wong-2019-stress-empatica, ollander-2016-stress-empatica, gjoreski2016continuous}, commercial devices such as those from Fitbit \cite{Staff_2022}, Garmin \cite{garmin-watch}, and Whoop \cite{Kuzmowycz_2023} have recently introduced stress tracking and in-the-moment interventions prompted by the detection of \edit{physiological} responses indicative of potential stress. 

\edit{These new capabilities are being used to improve stress management in two ways. First, passively collected data are being used to create new visualizations for self-reflection that show stress arousal levels in different spatial and temporal contexts~\cite{sanches2010mind,kocielnik2015personalized,kocielnik2013smart,stepanovic2019designing,xue2019affectivewall}. Qualitative evaluations have shown that these visualizations lead to new insights with a potential for self-initiated behavioral changes~\cite{kocielnik2013smart}. The therapeutic potential of such visualizations, however, is yet to be established as statistically significant changes in self-reported stress pre- and post-study were not found.

Second, the passive detection of stress has been used to deliver interventions at stressful moments~\cite{howe2022design}. Although the reduction in self-reported stress has been observed before and after some of these momentary interventions, only a negligible study-long reduction in stress has been found thus far~\cite{howe2022design,tong2023just,lee2020toward}. We, however, note this is still an active area of exploration given that the passive detection of stress has emerged only recently and efficacious interventions can take decades to emerge, accepted, and widely used. Interestingly, participants in these studies wished the momentary interventions to match the specific source of their stress, i.e., stressor, at the current moment~\cite{howe2022design,tong2023just}.

\begin{figure}[t]
    \centering
    \includegraphics[width=0.95\textwidth]{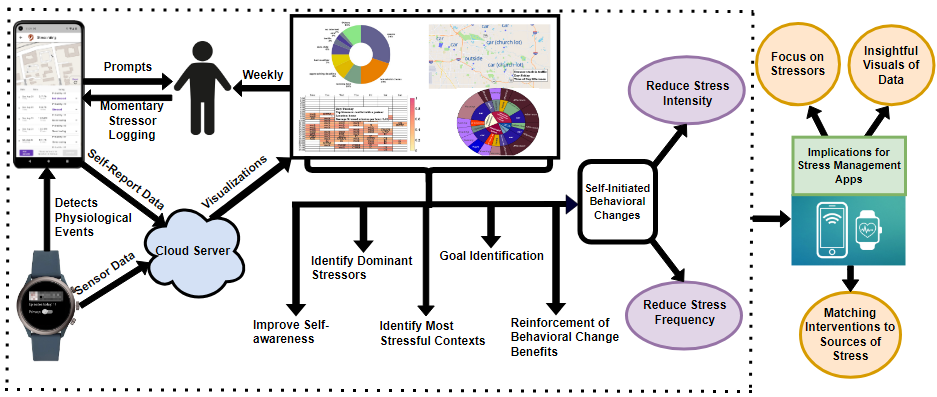}
    \caption{Key steps in the MOODS study along with impacts on study participants and implications for stress management apps.  }
    \label{fig:framework}
    \Description{The figure shows the overall process from stressor logging to self-reflection via visualizations leading to self-initiated behavior changes which had a positive impact on both stress intensity and frequency. The flow diagram shows how the event generation to stressor logging part was carried out which was later used on a weekly basis to generate insights via visualizations for participant's self-reflection. Stressor logging together with visualizations motivated participants to move toward or reinforce behavior changes in five different ways; improve self-awareness, identify dominant stressors, identify most stressful contexts, goal identification and reinforcement of behavioral change benefits.  It also shows how these led to self-initiated behavioral changes to have a positive impact on stress dynamics. Finally, it shows three implications this observational study can bring for stress management apps.}
\end{figure}

In this paper, we investigate the \emph{feasibility of collecting momentary stressors} and the \emph{utility of integrating stressors in self-reflective visualizations in reducing self-reported stress} in a longitudinal study.

We address several challenges in the presented study --- 1.) Although stress responses can now be detected automatically, most \emph{stressors can't yet be identified from sensor data}. How should stressors then be collected? 2.) If self-reporting is used to collect stressors, when should participants be asked to report stressors? Participants may not have the \emph{cognitive capacity} to engage in self-reporting at the time of stress~\cite{gross2015extended} and the recall of stressors usually \emph{fades from memory} over time~\cite{gilbert2009stumbling}. 3.) How should self-reports be collected so that we can capture the \emph{diversity} in stressors but \emph{minimize the participant burden}? 4.) How should \emph{self-reported stressors be integrated} into the visualizations of stress arousal data from sensors and the associated spatiotemporal contexts? 5.) How can visualizations provide \emph{new utilities} with every week of additional data in a longitudinal study so participants see a benefit in continuing to collect data and overcome the accruing \emph{burden, growing fatigue, and diminishing interest} over time observed in a longitudinal study?}

To conduct our research, we developed a new smartwatch app, a cross-platform smartphone app, and a companion cloud software. We deployed them in a 100-day field study completed by 122 participants. During the study, called Mobile Open Observation of Daily Stressors (MOODS), participants received prompts when \edit{physiological} events \edit{were detected} by the smartwatch app, \edit{indicated by changes in pulse physiology measured at the wrist level}. \edit{The prompts were delivered \emph{soon after the detected event concluded} so that the memory of the stressor would still be fresh, but they would have recovered from the recent event. Participants} were asked to rate their stress and log the stressor responsible (on the smartphone app) if they rated the event as stressful. \edit{We used \emph{predictive autocomplete} to limit self-report burden}. Each week, they received \edit{\emph{new types} of} visualizations \edit{that allowed them to dig deeper into} their data. \edit{To assess change in self-reported stress, the intensity of stress was measured via momentary ratings, and the frequency of stress was measured via weekly surveys. Qualitative feedback was collected during the exit interview to understand the mechanisms behind changes in self-reported stress. Participant burden was assessed from app usage logs; utility was measured from app approval ratings.} Importantly, no monetary incentives were provided, thereby emphasizing the intrinsic utility and burden of the app as the primary drivers of user retention. 

Figure~\ref{fig:framework} provides a visual summary of this work. Our key contributions are to 
\begin{enumerate}
    \item Show the feasibility of engaging participants to log their stressors upon the detection of a physiological event,
    \item Show that \edit{new visualizations of stress and stressors each week} can keep participants engaged, and
    \item Show that momentary logging of stressors and visualizations of stress and stressors data have the potential to increase self-awareness, trigger self-initiated behavioral changes, and decrease self-reported stress.
       
\end{enumerate}

\section{Related Works} \label{sec:related-works}
\edit{Our work investigates the feasibility of collecting self-reports of stressors triggered by sensor data, providing visualizations of stress for self-reflection, and assessing its impact on changes in self-reported stress. It builds upon and contributes to prior works on using sensor data to trigger self-report prompts, stressor logging, stress visualizations, and stress interventions. We discuss related work in each of these topics in the following.}

\subsection{Sensor-Triggered Prompting}\label{sec:related-works-intervention}

Real-time detection of changes in human states such as physical activity or transitions between activities from wearable devices has been used to improve the timing of prompting for self-reports or interventions~\cite{van2019effect,dunton2014development,aminikhanghahi2019context,akbari2021facilitating}. 

For example, in a study to assess physiological correlates of emotions~\cite{valence-arousal-variance,barrett2017emotions}, participants were asked in~\cite{lisa-context-aware-2020} to collect ECG, impedance cardiogram (ICG), and accelerometry data for 14 days. Whenever inter-beat interval (IBI) changed by a threshold, participants were asked to rate their emotions. Our work uses a similar study design but generates prompts based on the detection of \edit{physiological} responses, some of which are indicative of a potentially stressful event.

Furthermore, real-time detection of stress from physiological events ~\cite{sensor-stress-management2015,smith2020integrating,battalio2021sense2stop} or from digital traces such as emails~\cite{howe2022design} makes it possible to deliver interventions in moments of stress~\cite{sensor-stress-management2015,smith2020integrating,battalio2021sense2stop}. Fitbit Sense 2~\cite{Staff_2022} recently introduced a \edit{\emph{Body Response}} feature that detects when heart rate, heart rate variability, skin temperature, and electrodermal activity indicate a potential stress event. It then prompts participants to log their mood and engage in an intervention of their choice to manage their stress in the moment. Our approach is similar \edit{but we wait until the conclusion of the physiological event when participants are likely to have recovered from the event but are likely to still remember the stressor.}

\subsection{Logging and Journaling \edit{of Stressful Events} for Stress Management}
\edit{Stressor is an event or situation, whether social or emotional (e.g., challenge, threat, opportunity) that triggers a stress response, i.e., deviation from homeostasis~\cite{chrousos2009stress}. 
Successfully addressing stress usually requires activating coping mechanisms that directly address the responsible stressor~\cite{howe2022design,tong2023just}. Therefore, to guide the development of appropriate stress interventions, many works have focused on understanding prevalent stressors and their role in PTSD~\cite{silove1997anxiety}, depression~\cite{mazure1998life}, anxiety disorder~\cite{calancie2017exploring}, heart diseases~\cite{munzel2017environmental}, asthma or rheumatoid arthritis~\cite{vig2006role}, as well as in the general population~\cite{almeida2002daily,hartanto2023smartphone,klaiber2021ups}. These studies have shown that stressors predict daily symptoms and health outcomes. 

Stressors have been collected at the end of day, via telephones~\cite{almeida2002daily}, surveys~\cite{klaiber2021ups}, and smartphones~\cite{hartanto2023smartphone}. More frequent collections (5-6 times daily) have resulted in more stressors reported, and hence are being adopted more widely~\cite{zawadzki2019understanding}.

It has also been shown that writing about stressful events has been shown to improve health, psychological well-being, physiological functioning, and general functioning~\cite{ullrich2002journaling}.
This inspired many works on facilitating stressor journaling or even mood tracking to improve awareness. For example,} the DeepMood application~\cite{suhara2017deepmood} directed participants to input their moods and activities into a smartphone app three times daily, with the goal of anticipating episodes of depression. As participants actively engaged in the mood-monitoring process through these apps, they observed an increase in emotional self-awareness. \edit{Some works have even combined logging with stress intervention.} In~\cite{flinchbaugh2012student}, stress management techniques, gratitude journaling, a combination of both, and a control condition were investigated across four management course sections. The results indicated that students in the combined intervention and gratitude journaling groups exhibited increased levels of classroom engagement and perceived meaning.

\edit{As journaling involves a substantial participant burden, emerging technology has been increasingly leveraged to reduce participant burden~\cite{dubad2018systematic}.  In~\cite{hosseini2022multimodal}, data collected by a wearable was processed at the end of the day and detected stressful events were presented to the participants with an option to select from a list of 13 stressors to assist with the recall. Chatbots have been explored to engage participants in logging their moods for three weeks~\cite{kawasaki2020assessing}. To reduce the burden further, a single item was developed to be delivered on smartwatches so participants can log stress ratings with a quick glance and a tap~\cite{intille2016muema,king2019micro}.}

Some commercial devices have also adopted these approaches. For example, Fitbit Sense 2~\cite{Staff_2022} smartwatch asks participants to rate their mood and engage in an intervention such as reflecting on their feelings and triggers upon detecting stress-related physiological events. The WHOOP Journal~\cite{Kuzmowycz_2023} allows participants to document their perception of stress levels, enabling them to track and analyze how they subjectively experience stress \edit{in the smartphone app.

Logging of stressors still involves two types of burden --- the physical effort in entering the stressor and mental effort in recalling the stressor. We address both of these issues by prompting participants soon after the conclusion of a potentially stressful event to reduce the recall effort and by providing a predictive autocomplete to reduce the physical effort. We evaluate the former via weekly surveys and the latter via time taken to log the stressors from app usage logs.}

\subsection{Visualizations of Stress and Stressors} \label{sec:related-work-viz}

\edit{
A large body of work exists on presenting visualizations of stress data collected from wearables to facilitate new insights that can improve stress management and user engagement. The majority of these works combine stress arousal~\cite{sharmin2015visualization} (sometimes expressed as body response~\cite{Staff_2022}, bodily reactions~\cite{sanches2010mind}, arousal level~\cite{kocielnik2013smart}) with spatiotemporal context and self-reports~\cite{Staff_2022}. These visualizations have been targeted at self-reflection by participants~\cite{sanches2010mind,kocielnik2015personalized,kocielnik2013smart,stepanovic2019designing,xue2019affectivewall} and sometimes for experts' review~\cite{sharmin2015visualization}. In many cases, the visualizations are provided at the end of study~\cite{bari-2020-stressful-conversation,kocielnik2013smart,mcduff2012affectaura}, and in some cases, the participants had access to the visualizations throughout the data collection period~\cite{kocielnik2015personalized,sanches2010mind,xue2019affectivewall,jiang2023intimasea}. Most works targeted the visualization for private use by the participant~\cite{sanches2010mind,kocielnik2015personalized,kocielnik2013smart}, while some allowed visualization of data aggregated across multiple participants to be displayed in public places for reflection by the entire group or organization~\cite{xue2019affectivewall}. In~\cite{bari-2020-stressful-conversation,jiang2023intimasea}, partners were allowed to view the visualizations of each others' data together to understand the impact of each other's actions on the stress level of their partner. These works have shown the potential of stress visualizations to increase awareness, gain new insights, and improve stress management by promoting behavior change. 

Prior works also introduced a wide variety of templates. They include calendar view~\cite{kocielnik2015personalized}, line, bar, and box plot~\cite{stepanovic2019designing}, and pie chart~\cite{kocielnik2013smart}. We use the color coding of arousal levels obtained from sensors overlaid on calendars and bar charts to integrate spatiotemporal contexts with stress data from wearables, line graphs to show trends, box plots to show variability, and pie charts to show prevalence. We contribute to the body of works on stress visualization by proposing ways to integrate self-reported stressors with stress arousal and spatiotemporal contexts obtained passively. We target eight insights~\cite{choe2015characterizing} and introduce a new protocol, i.e., new visualization types each week (for 14 weeks) that allows participants to gradually dig deeper as they collect more data each week. 
}

\subsection{Stress Interventions} \label{sec:stress-intervention}
As we are evaluating the impact of our \edit{visualizations} on stress \edit{reduction}, we review some recent works on stress intervention to \edit{report} the magnitude of the study-long reduction in stress. \edit{Interventions can have a momentary reduction in stress as well as a reduction in study-long stress, with the latter indicating the potential of the intervention to have a lasting impact.} A recent work~\cite{howe2022design} delivered \edit{both} pre-scheduled interventions and stress-triggered interventions. They detected stress using email, calendar, camera, and heart rate. A chatbot prompted participants to practice one of three stress intervention exercises based on Cognitive Behavioral Therapy (CBT) and Dialectical Behavioral Therapy (DBT). In four weeks of using interventions, ($n=86$) participants reported approximately 6.8\% reduction in momentary stress post-intervention compared with pre-intervention (from 2.16 to 1.82 on a 5-point scale). However, no \edit{significant} study-long reduction in overall stress was \edit{found}.

In~\cite{tong2023just}, 30 participants used a browser-based application to receive stress micro-interventions for four weeks. One experimental sub-group reported the highest reduction of 14.8\% (mean reduction of 0.74 on a 5-point scale) in momentary stress between pre and post-intervention. However, \edit{no} overall reduction in study-long stress was \edit{found} in this study \edit{either}. In a 30-day study with 47 participants~\cite{lee2020toward}, a calendar-mediated stressor anticipation application was deployed. It allowed participants to anticipate expected stressful events. Participants who executed a problem-focused intervention reported a 5.8\% (-0.29 on a 5-point scale) reduction in stress between pre and post-event. Again, no study-long change in stress was found. A 28-day study with 65 participants~\cite{konrad2015finding} deployed a system that set adaptive goals in three coaching dimensions (Exercise, Meditation, and Accessibility). They reported a 22.5\% study-long reduction on one stress measure but did not find any significant reductions on two other stress measures. A study~\cite{suyi2017effectiveness} with 37 mental health professionals who performed a 6-week meditation-based stress reduction (MBSR) program, reported a 7.6\% reduction in study-long stress. In summary, study-long reduction in self-reported stress is usually low or negligible.

\edit{In contrast, our work investigates whether providing visualizations of self-reported stressors together with stress and associated spatiotemporal contexts as a personal informatics system~\cite{li2010stage}
can quantitatively reduce self-reported stress and keep participants engaged in a longitudinal study.}

\begin{figure}[t]
     \centering
     \begin{subfigure}[b]{0.33\textwidth}
         \centering
         \includegraphics[width=\textwidth]{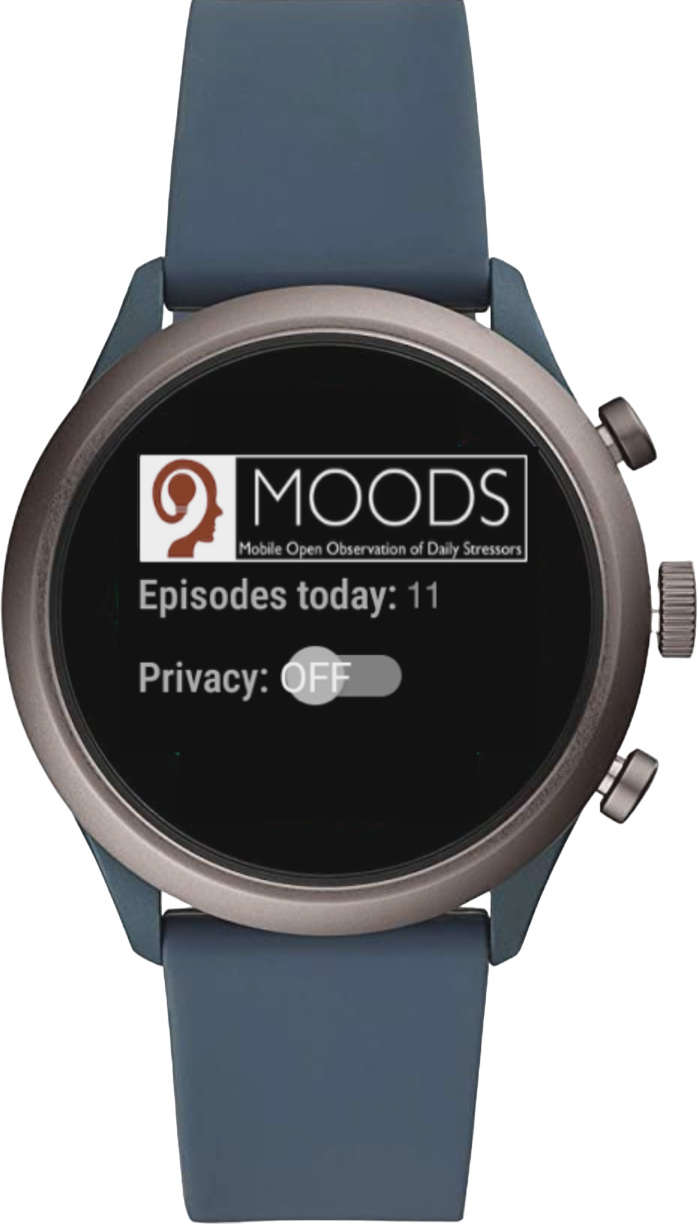}
         \caption{Smartwatch app home screen}
         \label{fig:watch_screen}
       
     \end{subfigure}
  \hfill
     \begin{subfigure}[b]{0.3\textwidth}
         \centering
         \includegraphics[width=\textwidth]{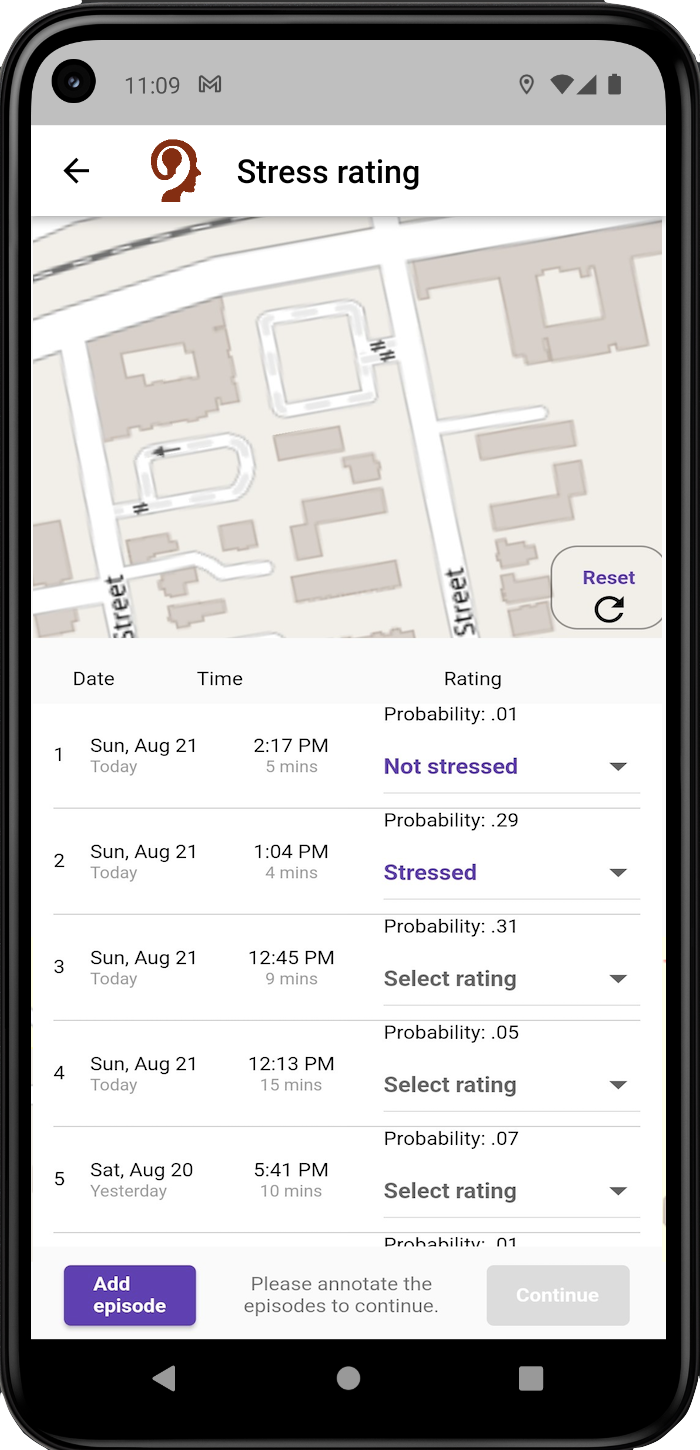}
         \caption{Smartphone app home screen}
         \label{fig:app-home-screen}
      
     \end{subfigure}
     \hfill
     \begin{subfigure}[b]{0.3\textwidth}
         \centering
         \includegraphics[width=\textwidth]{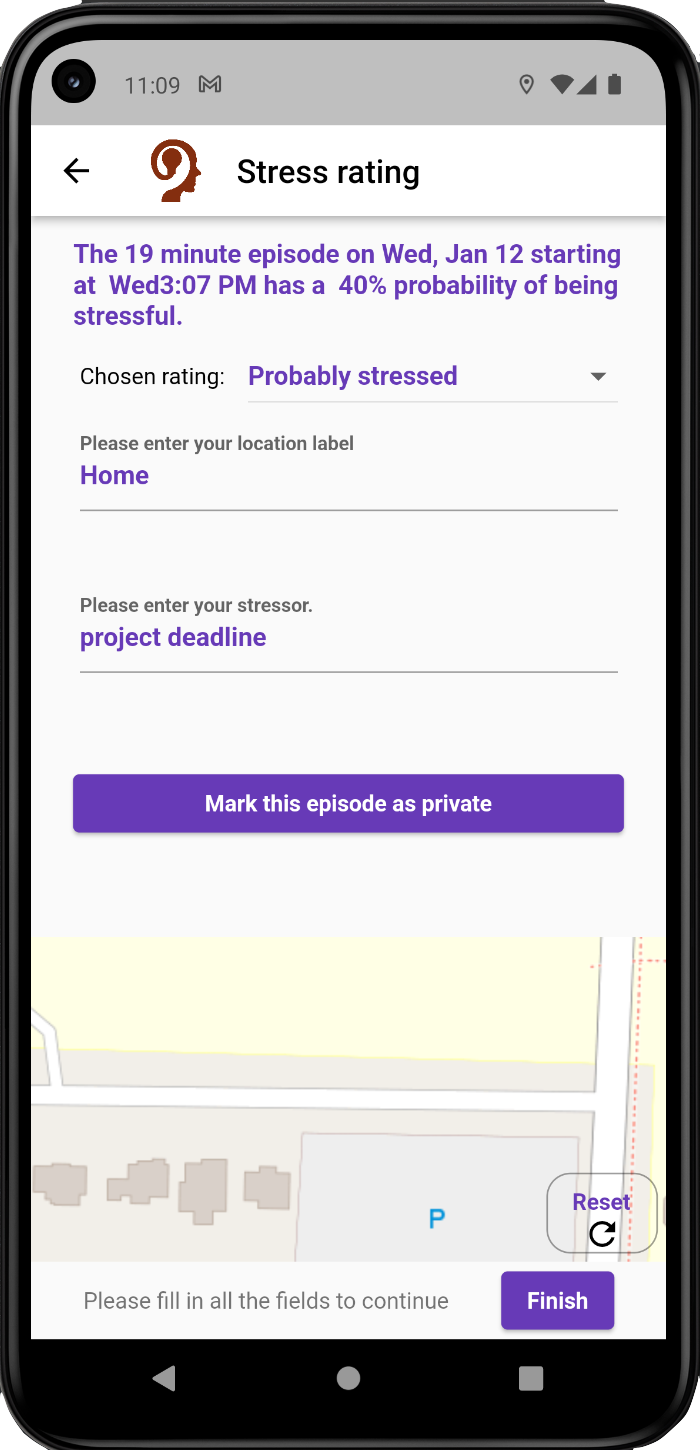}
         \caption{Stressor annotation screen}
         \label{fig:app-annotation-screen}
      
     \end{subfigure}
        \caption{\anonymize{MOODS} app screenshots for events on smartwatch and stressor annotation, review, and edits on smartphone }
        \label{fig:app-screens}
        \Description{This figure contains three subfigures—The first subfigure shows the MOODS study app watch interface, which shows the number of events detected in a day. The participant interface also gives participants the option to suspend data collection for an hour.

The second sub-figure shows the smartphone app screen which depicts all selected sensor-detected stress events and the corresponding stress likelihood of that event. Additionally, it shows the location of those events in a map interface.

The third sub-figure shows the smartphone app screen where participants can provide stressors and semantic locations for the stressed events. This page also shows the approximate location on the map where the stress event is detected.
}
\end{figure}

\section{Methods}

\subsection{Design Rationale for the MOODS Study}
Our apps \edit{(Figure~\ref{fig:framework})} were designed to serve as \edit{a Stage-Based Model of }Personal Informatics System~\cite{li2010stage}. Except for the \emph{Preparation} stage when participants select what data they want to collect and using which tool (e.g., deciding to participate in our study), \edit{the MOODS Study} aimed at helping participants with the remaining four stages. Many momentary stressors fade from people's memory soon after they are over unless it is a major event with a longer-lasting impact~\cite{gilbert2009stumbling}. Many of these stressors may not be reported when asked later on. Our \edit{study} was designed to help participants with the \emph{Collection} stage by prompting them to easily log their stressors soon after their occurrence while limiting burden, as suggested in~\cite{lee2020toward}. It helped them with the \emph{Integration} stage by summarizing their data in visual forms. To help them with the \emph{Reflection} stage, they were given the opportunity to review their stressor logs anytime and were presented with 16 different visualizations. These visualizations were meant to help them identify and execute specific \emph{Actions} to better manage their stress \edit{as well as to observe the impact of their self-initiated actions on their stress and stressors}.

\begin{figure}[t]
    \centering
    \includegraphics[width=0.7\textwidth]{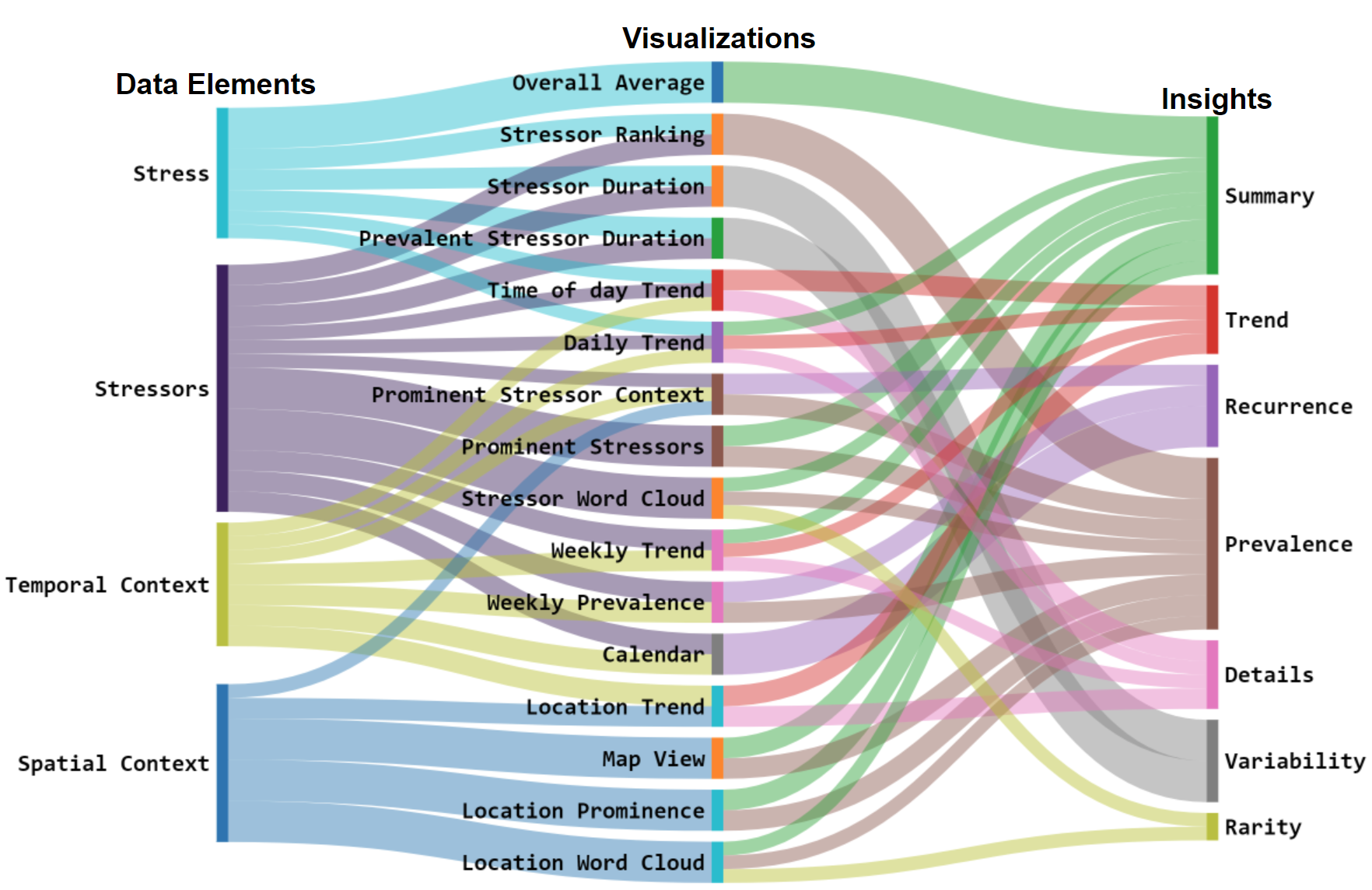}
    \caption{\edit{Overall design of weekly visualizations, showing which visualizations (in the center column) used which data elements (left column) and aimed for what insights~\cite{choe2015characterizing} (right column).}}
    \label{fig:vis_goals}
    \Description{
The figure shows the overarching design of weekly visualizations. 16 visualizations are shown in the center column of the figure. The data elements (stress, stressors, temporal and spatial context) based on which different visualizations were created, are shown in the left column. The rightmost column showcases the insights facilitated by different visualizations. There were total eight insights; summary, trend, recurrence, prevalence, details,variability and rarity. The figure is created using Sankey Template where visualizations in the center column are connected to corresponding data elements and the insights.}
\end{figure}
 
\subsection{MOODS System Design}
 
We developed an app for WearOS smartwatches, a cross-platform companion smartphone app, and cloud services.  


\subsubsection{Physiological Events, Stress Ratings, and Stressor Annotations}\label{sec:stressor-annotation}

\edit{Similar to recent works that use commercial smartwatches running pre-trained proprietary models~\cite{jiang2023intimasea, booth2022toward, ding2021data, akbar2021physician}, we use a pre-trained model (used in CuesHub smartwatch app~\cite{CuesHub})  
to detect \edit{physiological} events, indicative of physiological activation at the wrist level. Each event has a start time, end time, and duration. Similar to Garmin watches, the model we use processes wrist motion and Photoplethysmography (PPG) sensor data and produces a score between 0 and 100 that indicates the likelihood with which the current physiological event is considered to exhibit a stress response. 

To limit participant burden, we selected only a subset of physiological events for prompting. We gave a higher weight to high scores as a key focus of our study was to collect momentary stressors. Specifically, all events with scores $> 95^\text{th}$ percentile were selected, whereas 80\% of events in the $\{75^\text{th}, 95^\text{th}\}$, 10\% of events in the $\{25^\text{th}, 75^\text{th}\}$, and 20\% of events in the $\{0, 25^\text{th}\}$ percentile were randomly selected for an average of 3, 2, and 1 prompts per day, respectively. Our prompts cover all score ranges to obtain user-labeled sensor data in both stress and non-stress situations in real life.}

\edit{Participants received a prompt on both their smartwatch and their smartphone.} To determine which of the presented events were indeed stressful, participants were asked to choose a rating from the following Likert scale (\emph{`Not stressed,' `Probably not stressed,' `Unsure,' `Probably stressed,' `Stressed'}). If the annotation was \emph{`Stressed,' `Probably stressed,' or `Unsure'}, the smartphone app prompted them to provide the semantic location and describe the likely stressor (see Figure~\ref{fig:app-annotation-screen}). Using a Likert Scale allowed us to also understand the intensity with which participants perceive different momentary stressors. To reduce the burden, we incorporated a predictive text input module that displayed closest-matching stressors as participants typed in the input box. We prepopulated our database with 80 unique stressors \edit{(e.g., \texttt{traffic/transportation, work overload/demand, miscommunication, financial issues, interaction with boss, etc.})} from the literature~\cite{almeida2002daily}. This database was augmented, locally on the smartphone, with each newly-entered stressor to help participants annotate their specific stressors in a quick and consistent manner. To further assist them in recalling their stressors, participants were also shown the event location on a map along with the date, start time, duration, and the likelihood with which the app determined this event to be stressful (see Figure~\ref{fig:app-annotation-screen}). 

Once participants completed their annotations, they returned to the stressor dashboard, which also served as the smartphone app's home screen (see Figure~\ref{fig:app-home-screen}). The dashboard displayed all previously logged events with the option to edit the stress rating, stressor description, or to mark the event as private, which excluded those events from any analysis. Finally, the dashboard allowed participants to self-report a stress event manually.

\subsection{Weekly Visualizations \edit{to Increase Self-Awareness} of Stress, Stressors, \edit{and their Surrounding Contexts}}
We designed 16 different visualizations from the stress ratings and stressor logs of participants. The goal of these visualizations was to help them self-reflect and gain \edit{1.) better realization of stressful moments, 2.) better awareness of momentary stressors, and 3.) a better understanding of spatio-temporal contexts surrounding stressors.} 

\subsubsection{Design Rationales for Weekly Visualizations}
\edit{Figure~\ref{fig:vis_goals} shows the overall design of our visualizations. They used the following data elements --- 1.) Stress (ratings and duration), 2.) Stressors, 3.) Temporal Context (time of day, day of week), and 4.) Spatial Context (places and locations). They aimed to encode patterns over these data to facilitate the following insights --- 1.) Summary, 2.) Trends, 3.) Recurrences across time and place, 4.) Prevalence, 5.) Details, 6.) Variability,  7.) Rarity, and 8.) Comparison (facilitated via interactivity in the visualizations). These insights are based on the eight categories of insights proposed for self-reflection during the \emph{Reflection} and \emph{Maintenance} stages of personal informatics~\cite{li2010stage} among quantified selfers~\cite{choe2015characterizing}. We made two modifications --- 'Prevalence' was used for 'Self-reflection' to discover dominant stressors and 'Recurrence' was used for 'Correlation' to discover associations of stressors with routine, place, or specific daily behavior. Figure~\ref{fig:vis_goals} is a Sankey Diagram where  \emph{Data elements} and \emph{Insights} are two types of source nodes and visualizations are the target nodes. Among \emph{Data element} nodes, 'stressors' has the largest number of links (11), while 'prevalence' and 'summary' have the largest number of links among \emph{Insight} nodes with 8 links.

Given a large number of combinations of data elements and insights ($4\times 8$) and diversity among them, we selected templates that suited each purpose. For example, for the overall summary, we used an angular gauge chart (as in fuel gauge indicators), sunburst chart to show the relationship among top stressors and their spatial and temporal context, a donut chart to show prevalence, a calendar overlay to show recurrence, line charts for trend, bubble chart to combine prevalence and trend, scatter plot for combining details with trend and prevalence, a combination of violin and box plot to show variability, and word cloud to combine prevalence with rarity.
}
Excess abstractions in visualization sometimes hinder participants from gaining valuable insights from their data~\cite{rapp2016personal}. Hence, we utilized participants' own wording in most visualizations, making only minor spelling corrections. Where the input text was lengthy, we utilized abbreviations formed from the initial letters of each word to condense the information for visualization purposes.

\begin{figure}[t]
     \centering
     \begin{subfigure}[b]{0.45\textwidth}
         \centering
         \includegraphics[width=\textwidth]{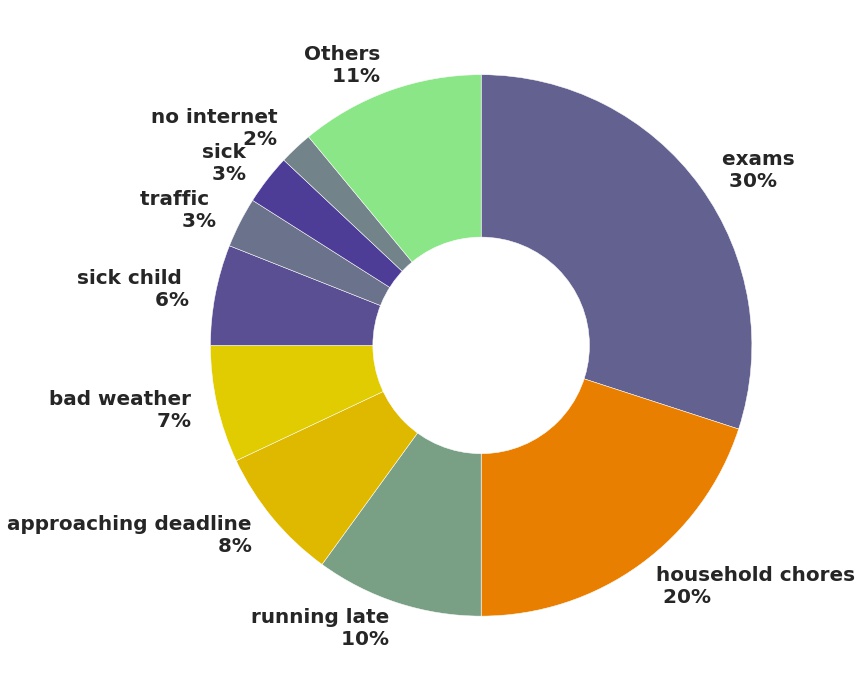}
         
         \captionsetup{justification=centering}
         \caption{Prominent Stressors }         \label{fig:Stressor_prm}
     \end{subfigure}
     \hfill
      \begin{subfigure}[b]{0.45\textwidth}
         \centering
         \includegraphics[width=\textwidth,height=5cm]{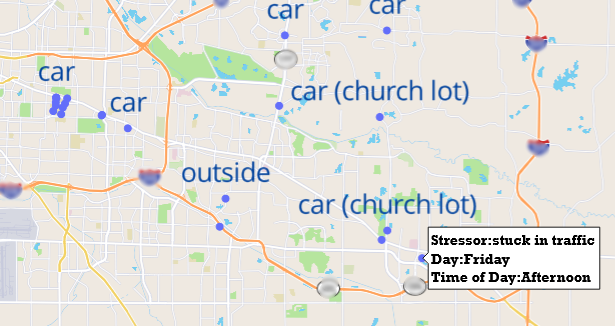}
         \captionsetup{justification=centering}
         \caption{Map View }
         \label{fig:map}
     \end{subfigure}
     \hfill
     \begin{subfigure}[b]{0.49\textwidth}
         \centering
         \includegraphics[width=8 cm,height=6 cm]
         {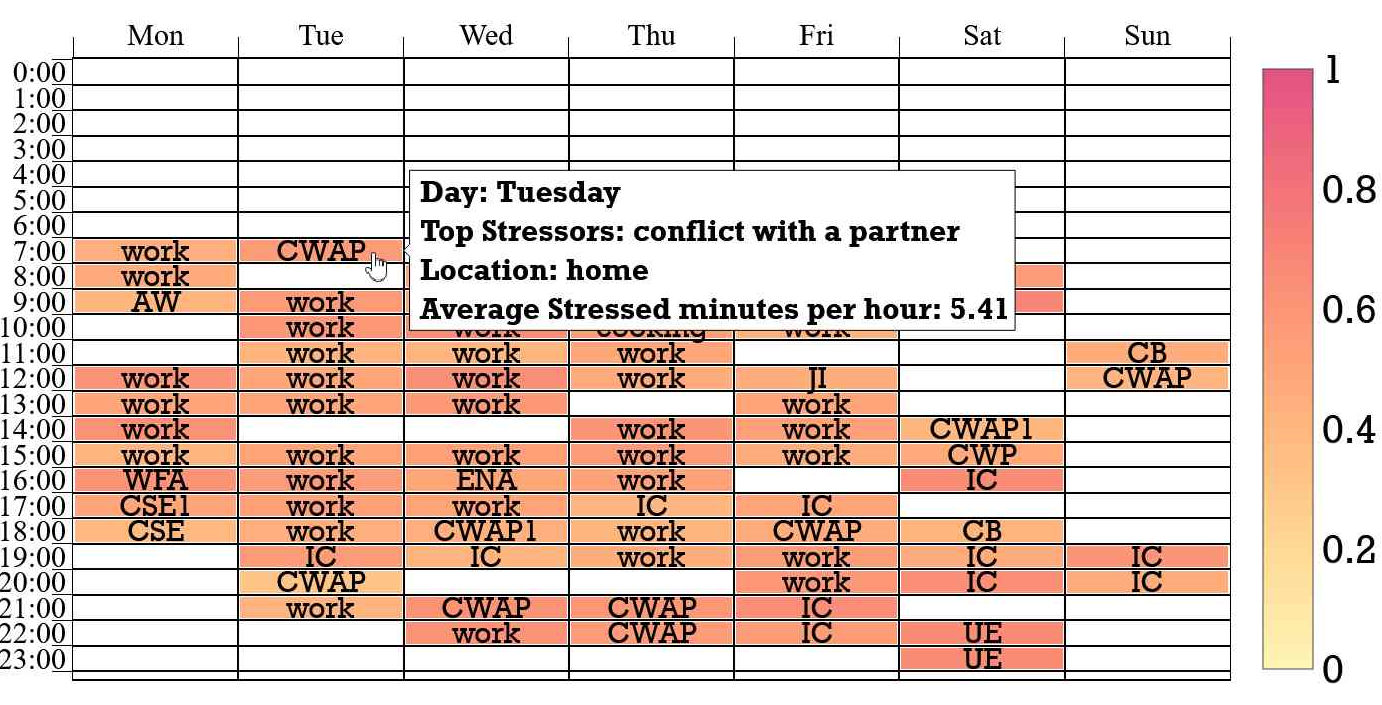}
         \captionsetup{justification=centering}
         \caption{Calendar View }
         \label{fig:calendar}
     \end{subfigure}
     \hfill
     \begin{subfigure}[b]{0.4\textwidth}
         \centering
         \includegraphics[width=\textwidth]{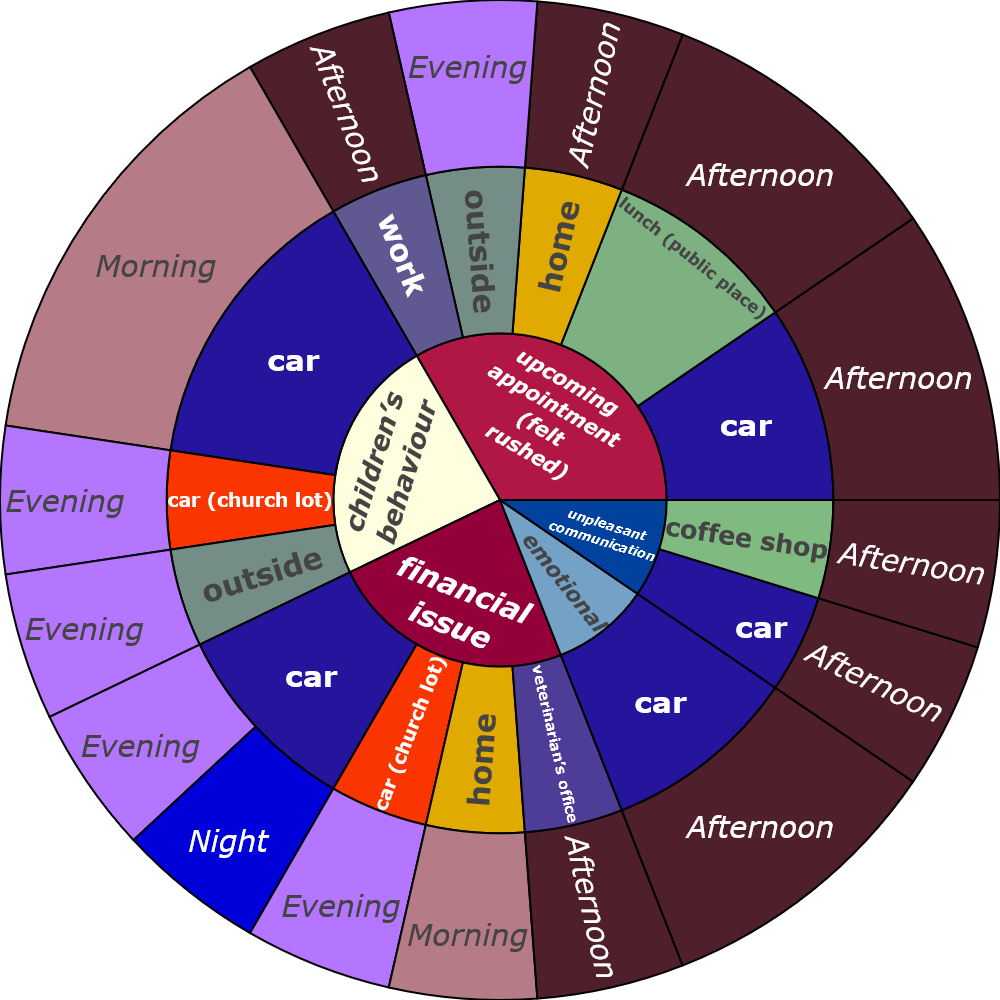}
         \captionsetup{justification=centering}
         \caption{Prominent Stressor Context }
         \label{fig:prominent}
     \end{subfigure}
        \caption{Examples of 4 visualizations. Details on each visualization can be found in Supplementary materials.}
        \label{fig:visualizations}
        \Description{  The figure shows four examples of visualizations. The first figure called Prominent Stressor shows the distribution of duration of most prevalent stressors. The second figure called Map View overlays the GPS coordinates of the stressed locations on a map.  The calendar view pinpoints the specific days of the week and times when they were most impacted. Prominent Stressor Context shows the location and time of day for the most prevalent stressors.

}
\end{figure}

\edit{All except three visualizations were} interactive. Participants had the option to tap on any data point to unveil the complete form of abbreviations and access additional details, such as location, time of day, or day of the week associated with the stressors as shown in Figures~\ref{fig:map} and~\ref{fig:calendar}. Moreover, the colors were selected to be friendly for those with color blindness. Visualizations were generated using~\cite{plotly} and~\cite{Hunter:2007}. Finally, we ensured that the visualizations were responsive for viewing on various screen sizes including smartphones, tablets, and desktops or laptops. Figure~\ref{fig:visualizations} shows four of these visualizations. More details about each visualization appear in the supplementary materials.

\subsubsection{Chronological Order of Weekly Visualizations}
After completing the first week of data collection, participants received two visualizations. A new visualization was added in each subsequent week. By incorporating cumulative data into each visualization, participants could see how their stress patterns evolved over time.
\edit{The chronological ordering of the visualizations was largely driven by the sufficiency of data for each visualization to be meaningful. Hence, those providing an overall summary or top stressors were introduced early on, followed by those displaying prevalence, recurrence (e.g., calendar), trend, details of individual stressors (that are very sparse early on), and variability. Word clouds were saved for the end as they needed a large amount of data to display both prevalence and rarity. Figure~\ref{fig:vis_rank} shows the order in which the 16 visualizations were introduced in the MOODS study over 14 weeks.
}

\subsection{MOODS Study Description}

\subsubsection{Procedure}
We obtained \edit{Institutional Review Board (IRB)} approval for the MOODS study. Interested participants completed a brief screening survey via Qualtrics: Eligibility criteria included having United States residency, ability to freely and independently provide legal consent, ownership of a sufficiently modern smartphone (queried based on operating system version), willingness and ability to follow study instructions in the English language, regular and sufficient access to mobile internet service, understanding of shortened smartwatch battery life, and willingness to recharge the smartwatch throughout the day. 

Participants selected for enrollment were sent an email providing personalized instructions for installing the MOODS study app on their smartphones, ensuring only selected participants could proceed with the enrollment process. Once installed, participants reviewed and completed the informed consent form via the smartphone app; participants were also provided a hyperlinked (PDF) document so that they could read the consent form on another preferred device. Participants who downloaded the MOODS study app but did not consent were afforded no further app functionality; all participants were informed they could uninstall the app with no penalty or further obligation to the study.

Consented (enrolled) participants completed a brief stress awareness assessment via the smartphone app. Participants also provided their demographic information (i.e., age, gender, occupation) and their shipping address. We then shipped them a Fossil Sport smartwatch and the instructions to configure the smartwatch with their smartphone via the MOODS study app. 
Every week, participants were sent weekly visualizations of their stressor patterns via their study-provided email and asked to complete brief weekly questionnaires via the smartphone app. 

No ongoing or end-of-study incentive was provided, to simulate real-life usage. \edit{At the conclusion of the MOODS study, participants were allowed to retain the smartwatch. They were asked to collect data for a period of 100 days, however, they could exit the study at any time without returning the smartwatch.} Even though some participants continued using the apps for up to 173 days, we limited our analysis to 14 weeks of data to maintain consistency and alignment with the planned study duration. Upon completion, participants received an email that included an invitation to either complete an end-of-study interview (virtually) or complete a questionnaire (via Qualtrics). Any participant who completed 100 days of data collection or was deemed to have stopped participating after completing at least 30 days of data collection was invited for an exit interview. 

\edit{\subsubsection{Exit Interview} During the exit interview, we collected participant feedback covering various aspects, including their interactions with the MOODS study apps, experiences with visualizations, and any goals or behavioral changes initiated throughout the study. Participants shared details on their preferred hand for wearing the smartwatch, devices used for reviewing weekly visualizations, potential interference of study apps with daily activities, and their overall preferences and dislikes concerning the app. Ratings were gathered regarding the utility of weekly visualizations in understanding stress patterns, visualization preferences, and the reasons behind participant likes and dislikes. Participants were also asked about any self-improvement goals established before or during the MOODS study and whether those goals were achieved. Additionally, inquiries were made into behavioral changes implemented during the MOODS study, participants' awareness of stress and stressors, their confidence in sustaining behavioral changes, and the likelihood of seeking out devices or apps for continued stress monitoring. Notably, participants were not directed to set goals or make behavioral changes pre or during the study; any such initiatives were self-initiated. The goal was to assess whether stressor logging with self-reflection aided participants in achieving their goals and managing stress through behavioral changes.}

\subsubsection{Participants}
\edit{MOODS apps} were tested by 22 pilot participants for 12 weeks to debug and improve the apps and study procedures for long-term remote operation. For the main phase, we began our recruitment by sending IRB-approved email invitations to academic groups, posting messages in professional media (e.g., LinkedIn), and posting flyers in various organizations with the intention of obtaining a diverse participant pool across occupation, geographical, and demographic attributes. We sent enrollment emails to 302 participants out of which 183 enrolled via the MOODS study app, and 143 were able to successfully configure the smartwatch with the smartphone. Out of 136 participants who began collecting data, we report data from $n=122$ (39 men, 77 women, and 6 non-binary) who annotated at least one stressor. They come from 13 different U.S. states. Of these, 71 were professionals in occupations such as academia, information technology, administration, public safety, health, legal, sales, and construction. Fifty (50) were students pursuing higher education, and one participant did not disclose their occupation. The average age of participants was \edit{$38(SD= 13)$ years}. 
\edit{We use `$SD$' to denote one standard deviation.}
The majority of participants ($n=78$) were iPhone users.    
Among the 46 participants who participated in the exit interview, 7 chose Zoom, 3 opted for phone, and the remaining completed a Qualtrics survey. For the rest of the paper, we use $P\#$ to associate specific participants with their testimonials.

\subsection{Stress Metrics and Statistical Tests} \label{sec:data-processing-analysis} 
\subsubsection{Stress Metrics} \label{sec:stress-metrics}
Participants rated their stress multiple times daily when prompted. These stress ratings were converted to a scale of 0 to 4, with 0 representing \emph{Not Stressed} and 4 representing \emph{Stressed.} These ratings provide us an opportunity to analyze the intensity of those prompted events, which we call \emph{Stress Intensity}. Additionally, each Sunday morning, participants were asked to indicate the number of times they experienced stress in the past week each day, with four response options: \texttt{At most once}, \texttt{More than once but at most twice}, \texttt{More than twice but at most three times}, and \texttt{Four or more times}. These responses were mapped to numerical values ranging from 1 to 4. We use this value as a metric of  \emph{Stress Frequency}. We use weekly surveys instead of stressor logs for frequency as not all stress occurrences can be captured via prompts, e.g., when the watch is not worn, no prompts are issued.

\subsubsection{Statistical Tests} \label{sec:statistical-tests}
To determine trends in self-reported measures of stress intensity and frequency, we used the Mann-Kendall Trend test~\cite{Hussain2019pyMannKendall}, which is commonly employed for analyzing consistently increasing or decreasing trends in time series data. In the rest of the paper, we use MK test for the Mann-Kendall test, "$m$" to denote the coefficient of slope, and "$b$" for the coefficient of intercept. We used the linear mixed effects model (LMM) for regression analyses of stress intensity and frequency with \edit{fixed effects (constant intercept and slope (with respect to week) across all participants) and two random effects for each participant; a random intercept and a random slope (with respect to week)}. To assess the differences within or between groups of participants, first, the Shapiro-Wilk test was performed to check for the normality of the data. Since most data from this study did not fit a normal distribution, we used the non-parametric Mann-Whitney U test for independent samples and the Wilcoxon signed-rank test for dependent samples for statistical comparisons. We used Python and R to process the data and perform the statistical analyses.

\section{Results}
\edit{We start by analyzing participant retention and data contribution to demonstrate the feasibility of sensor-prompted logging of stressors in a longitudinal study. Next, we present the like and dislike mentions for comparative evaluation of the 16 visualizations. We then statistically evaluate the effects of momentary stressor logging and visualizations on stress intensity and frequency. To understand the mechanisms behind a reduction in self-reported stress, we analyze increases in self-awareness, initiation of behavior changes, and its impact on stress. Finally, we analyze the physical and mental effort involved in logging stressors and the approval ratings of the MOODS study apps to assess their utility.}

\subsection{\edit{Participant Retention and Data Collection in the MOODS Study}}
\label{sec:user-retention-contribution} 
\begin{figure}[t]
     \centering
     \begin{subfigure}[b]{0.6\textwidth}
         \centering
         \includegraphics[width=\textwidth]{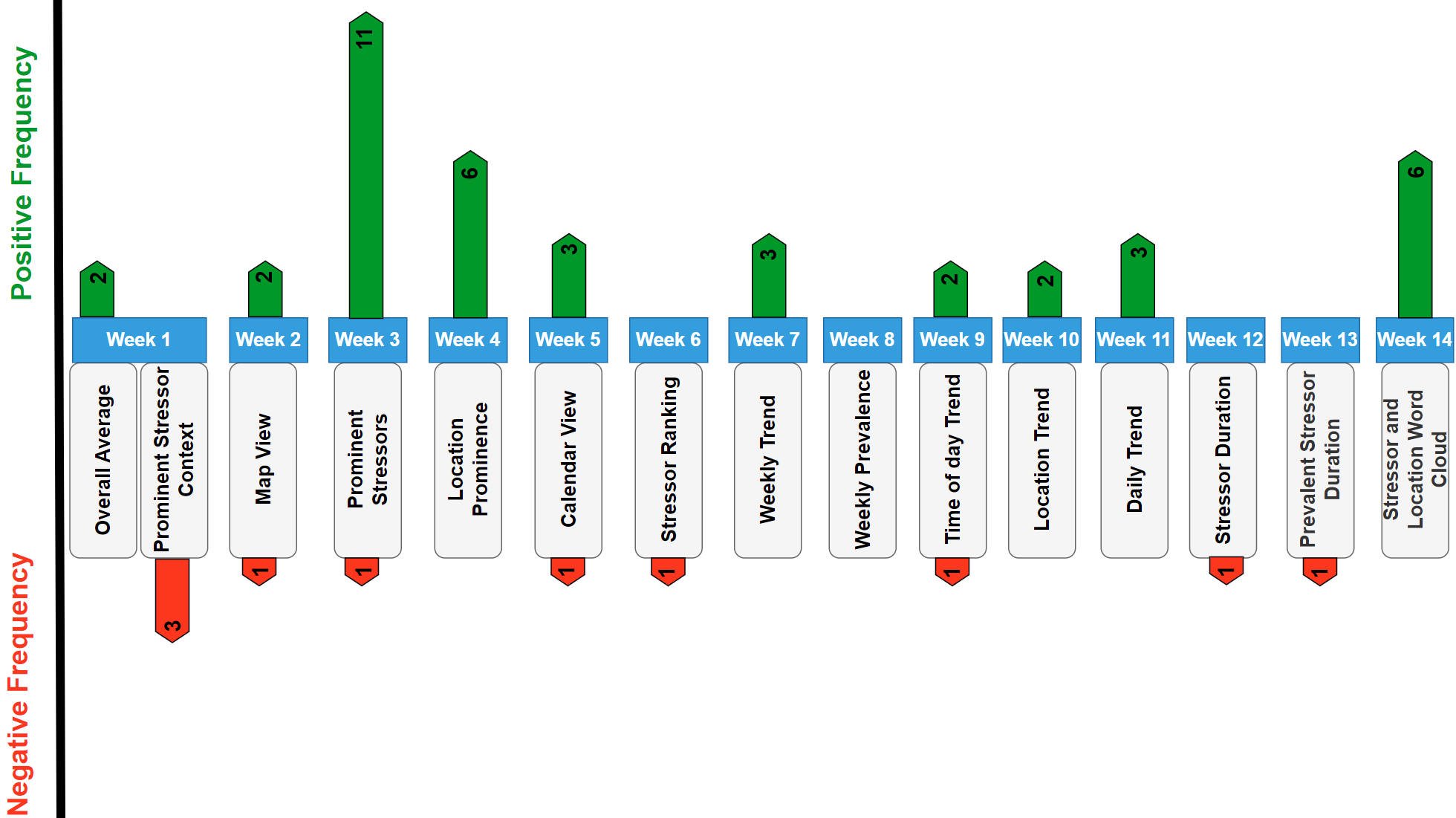}
         
         \captionsetup{justification=centering}
        \caption{\edit{Chronological sequence of weeekly visualizations (blue box) and their frequency of positive and negative mentions in participants' exit interviews}}
    \label{fig:vis_rank}
     \end{subfigure}
     \hfill
     \begin{subfigure}[b]{0.35\textwidth}
         \centering
         \includegraphics[width=\textwidth,height=5.5 cm]{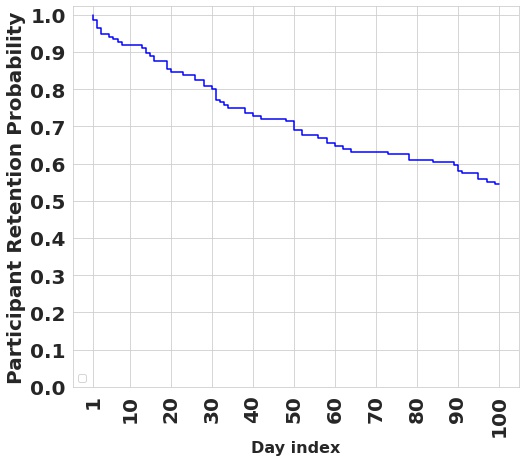}
         \captionsetup{justification=centering}
        \caption{Participant retention rate}         \label{fig:survival}
     \end{subfigure}
\hfill
\caption{\edit{(a) Chronological sequence of weeekly visualizations (blue box) and their frequency of positive and negative mentions in participants' exit interviews and (b) Participant retention}}
    \label{fig:vis_rank_retention}
    \Description{The first figure depicts the arrangement of visualizations based on their weekly delivery (indicated by the blue box). Additionally, it illustrates the frequency of positive mentions (represented by green bars) and negative mentions (indicated by red bars) as reported by participants during the exit interview. Prominent Stressors had the most positive mentions at 11 followed by Location Prominence and Word Clouds with 6. Similarly, Prominent Stressor Context had the most negative mentions with 3.
    
    The second figure shows the survival plot of participants in the MOODS study. Out of 136 participants who began collecting data, 110 remained active in the study for at least 30 days, resulting in a 30-day retention rate of 81\%. }
\end{figure}

\edit{To understand how engaging the sensor-prompted stressor logging and weekly visualizations were for the study participants, we analyze the participant retention in our study. Figure~\ref{fig:survival} shows the survival plot depicting participant retention in our study. Out of 136 participants who began collecting data, 110 remained active in the study for at least 30 days, resulting in a 30-day retention rate of 81\%.

On days with at least one prompt, participants} received an average of \edit{5.2} prompts per day, for a total of 35,981 \edit{prompts}. Participants responded to \edit{3.86} (i.e., 74\%) prompts per day, resulting in 26,732 stress ratings. Participants marked 211 events as private. An average of \edit{1.62} stressors were reported per day, resulting in 11,222 stressors \edit{that consisted of 1,476 unique stressors}. On average, each participant logged \edit{$83(SD= 76.5)$} stressors and an average of \edit{$21(SD= 17)$} unique stressors. \edit{The ten most reported stressors were \emph{anxiety}, \emph{approaching deadlines}, \emph{heavy traffic}, \emph{too much work}, \emph{acute or chronic pain}, \emph{work}, \emph{unsure}, \emph{playing video game}, \emph{research}, and \emph{unpleasant conversation}.} A total of 1,057 weekly surveys were completed. 

\subsection{\edit{Comparative Evaluation of} Weekly Visualizations} \label{sec:viz-evaluation}
In their weekly survey, participants were asked what impact the most recent visualization had on them. As new visualizations were introduced each week, we could potentially use their responses for evaluating each visualization. However, the visualizations were cumulative (for the benefit of participants) in that all previously introduced visualizations gradually became more useful as they showed the cumulative data, presenting ambiguity in using this method of evaluation. We, therefore, use their exit interview responses. Participants were asked about their preferences regarding the visualizations used in the MOODS study. They were given an open-ended question to express which visualizations they liked the most and the least. We examined the frequency of mentions for each visualization and categorized them as positive or negative. Figure~\ref{fig:vis_rank} shows the frequency of these mentions for each visualization \edit{(green and red bars indicate positive and negative mentions respectively). Four participants mentioned that they liked all visualizations without mentioning a specific one. Hence, their responses are not included in the positive mentions of any visualization.}

We now present some participant comments expressing why they like some visualizations more than others. Participants \edit{preferred visualizations that were \emph{easier to understand}.} $P41$ said that \emph{"The map view \edit{(Figure~\ref{fig:map})}, prominent stressors \edit{(Figure~\ref{fig:Stressor_prm})}, location prominence, and calendar \edit{(Figure~\ref{fig:calendar})} were visually easy to understand. The prominent stressor context \edit{(Figure~\ref{fig:prominent})} was difficult to understand."} $P4$ mentioned that \emph{"I liked the word cloud, as I'm a language person and it was visual and easy to understand. The prominent stressors graph was helpful."} 

\edit{Another factor was related to \emph{readability}. $P26$ said \emph{"I really liked word cloud, especially bold for prevalent stressors."} \emph{Readability} was mentioned by other participants, especially when it presented a challenge. $P27$ said \emph{"Schedule grid hard was to read for calendar view."}} $P10$ said \emph{"I did not like the prominent stressors graph because many of the acronyms did not make sense, and I did not know what it was referring to."} \edit{Similarly, $P16$ said \emph{"Sunburst chart (Prominent Stressor Context) was difficult because of small text."} }

\edit{A third factor cited related to the use of colors. $P8$ said \emph{"I noticed patterns across time of the day. Example: trouble sleeping (stressor). Chart with the different colors helped with that."}

Although there were some common themes that made some visualizations more or less liked by many participants, offering a choice of 16 visualizations that showed different aspects of their data and aimed at eight different insights helped many participants (with differing needs and preferences) find something they could benefit from. One participant ($P28$) even mentioned explicitly, \emph{"It was awesome to have so many options for visualizations presented."}}

\subsection{\edit{Reduction in Study-long} Self-reported Stress}

\begin{figure}[t]
     \centering
     \begin{subfigure}[b]{0.46\textwidth}
         \centering
         \includegraphics[width=\textwidth]{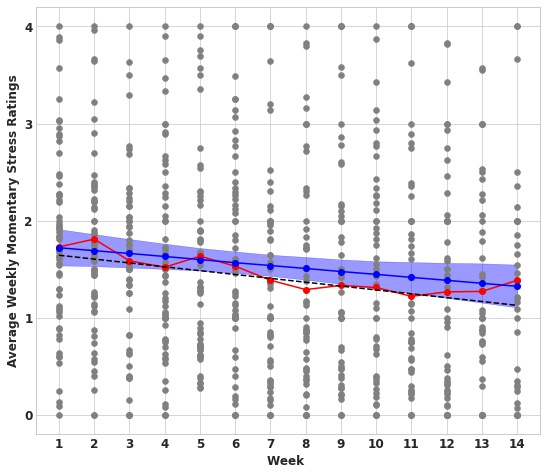}
         
         \captionsetup{justification=centering}
         \caption{Average weekly stress intensity}         \label{fig:avg_ratings_momentary}
     \end{subfigure}
     \hfill
     \begin{subfigure}[b]{0.46\textwidth}
         \centering
         \includegraphics[width=\textwidth]{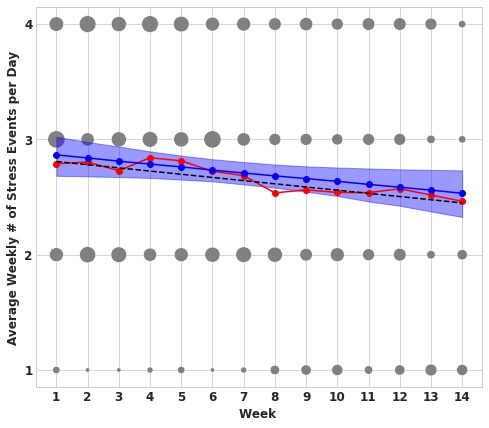}
         \captionsetup{justification=centering}
         \caption{Average weekly stress frequency}
         \label{fig:avg_ratings_survey}
     \end{subfigure}
\hfill

        \caption{Impact of self-reflection on the intensity and frequency of self-reported stress during the 14-week study. Red dots represent the population average for each week, the dashed lines represent the trend line, and the band around the blue line (population level LMM fit) represents the $5^\text{th}$ and $95^\text{th}$ percentiles calculated via bootstrap using resampling and replacement.}
        \label{fig:weekly_stress_ratings}
        \Description{The first figure shows the weekly Stress intensity average for different participants over the 14-week study. The red dots represent the average value of Stress intensity across all participants for each week. The size of grey dots in the second figure represent the proportion of participants who selected a specific value of Stress frequency in their weekly response for that week. The red dots represent the average value of Stress frequency across all participants for each week. The dashed lines in each figure represent the trend line, and the band around the blue line (population level LMM fit) represents the 5th and 95th percentiles calculated via bootstrap using resampling and replacement. To assess whether the weekly averages of stress intensity and frequency in both figures show an increasing or decreasing trend over the 14-week period, we used the MK test which showed that there is a statistically significant decline
in the trajectory (depicted by dashed lines) for both stress intensity  and
stress frequency. The estimated study-long decrease over the 14-week period in stress intensity and stress frequency are about 11\% and 9.5\% respectively.
}
\end{figure}

We computed weekly averages to measure the \emph{Stress intensity} for the past week for each participant. Grey dots in Figure~\ref{fig:avg_ratings_momentary} represent the weekly \emph{Stress intensity} average for different participants over the 14-week study. The red dots represent the average value of \emph{Stress intensity} across all participants for each week.
The size of grey dots in Figure~\ref{fig:avg_ratings_survey} represents the proportion of participants who selected a specific value of \emph{Stress frequency} in their weekly response for that week. The red dots represent the average value of \emph{Stress frequency} across all participants for each week.

To assess whether the weekly averages of \emph{stress intensity and frequency} (in Figures~\ref{fig:avg_ratings_momentary} and~\ref{fig:avg_ratings_survey}) show an increasing or decreasing trend over the 14-week period, we used the MK test. We find that there is a statistically significant decline in the trajectory (depicted by dashed lines) for both \emph{stress intensity} ($b=1.64$ and $m=-0.039$\oneS, $p < 0.05$) and \emph{stress frequency} ($b=2.81$ and $m=-0.027$\oneS, $p < 0.05$), where $b$ represents the intercept and $m$ the slope. The estimated study-long decrease over the 14-week period in \emph{stress intensity} and \emph{stress frequency} are about 11\% and 9.5\% respectively.  

We next check whether these decreases may be due to initial elevation bias and subsequent decline bias as has been observed in repeated measures~\cite{shrout2018initial}. Initial elevation bias refers to the tendency of participants to report higher levels or intensity of a target state in the initial survey (first few reports) than they actually experience, while decline bias refers to the tendency to report lower levels of intensity in subsequent reports (after the first few reports). In their study, the authors concluded that initial elevation bias is more pronounced than later decline bias for negative mental states in longitudinal studies of thoughts, feelings, and behaviors. To explore whether the decreasing trend in stress ratings we observed was influenced by initial elevation bias, we excluded the measurements from the entire first week for both \emph{stress intensity and frequency}. Even after this adjustment, the MK test still revealed a significantly decreasing trend. Therefore, we can conclude that the initial elevation bias did not account for the reductions observed here.

In the MK test, we used all data from all participants. However, there is wide variability between participants in self-reported stress. To check whether the trends persist when the self-reported stress measurements from each participant are analyzed separately, we employed linear mixed-effects models (LMM). We incorporated both fixed and random effects to account for the inherent participant variability. 
We excluded participants who had less than 5 weeks of survey ratings and participants who provided the same responses to all of their weekly surveys, i.e., had no variability. This left us with 63 participants. For \emph{stress intensity}, $b=1.76 (\pm 0.76)$ and $m=-0.03\oneS (\pm 0.062)$ ($p <0.05$) for fixed effects.  
For \emph{stress frequency}, $b=2.86 (\pm 0.74)$ and $m =-0.022\oneS (\pm 0.066)$ ($p <0.05$). Although decreasing trends were statistically significant for both \emph{stress intensity} and \emph{stress frequency}, they were not uniform across all participants, as the random effects were statistically significant for each measure. 

To see how the slope and intercept obtained in the LMM compare with that obtained via the MK test, we conducted the MK test for the same 63 participants that were analyzed in LMM. The MK test yields ($b=1.61$ and $m=-0.041\oneS$) ($p <0.05$) for \emph{stress intensity} and ($b=2.84$ and $m=-0.029\oneS$) ($p <0.05$) for \emph{stress frequency}, showing a similarity in both the slope and the intercept. We conclude that the participants reported a decreasing level of stress over time. 

\edit{To add some meaning to the magnitude of reduction in self-reported stress, we compare the data from the first week to that from Week 14. For \emph{stress intensity}, we see from the population-level fit (blue line) in Figure~\ref{fig:avg_ratings_momentary} that the average self-rating of momentary stress reduced from being close to \emph{Unsure} (i.e., 1.72) to \emph{Probably Not stressed} (i.e., 1.32).

For \emph{stress frequency}, the population-level fit (blue line) in Figure~\ref{fig:avg_ratings_survey} shows that the frequency of stress events shows a reduction of 0.33 (=2.86-2.53) per day, resulting in 10 fewer stress events per month.}

\begin{figure}[t]
     \centering
     \begin{subfigure}[b]{0.4\textwidth}
         \centering
         \includegraphics[width=\textwidth]{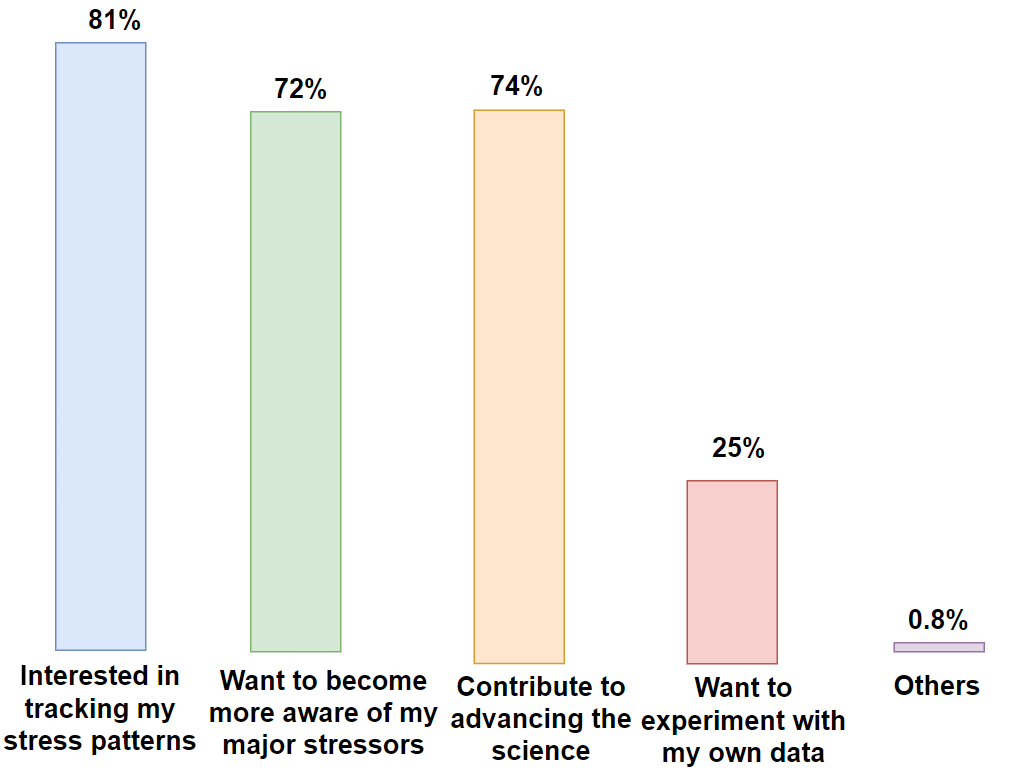}
         
         \captionsetup{justification=centering}
         \caption{\edit{Motivations reported for participating in the MOODS study in the Pre-Study survey}}         \label{fig:motivation}
     \end{subfigure}
     \hfill
     \begin{subfigure}[b]{0.55\textwidth}
         \centering
         \includegraphics[width=\textwidth]{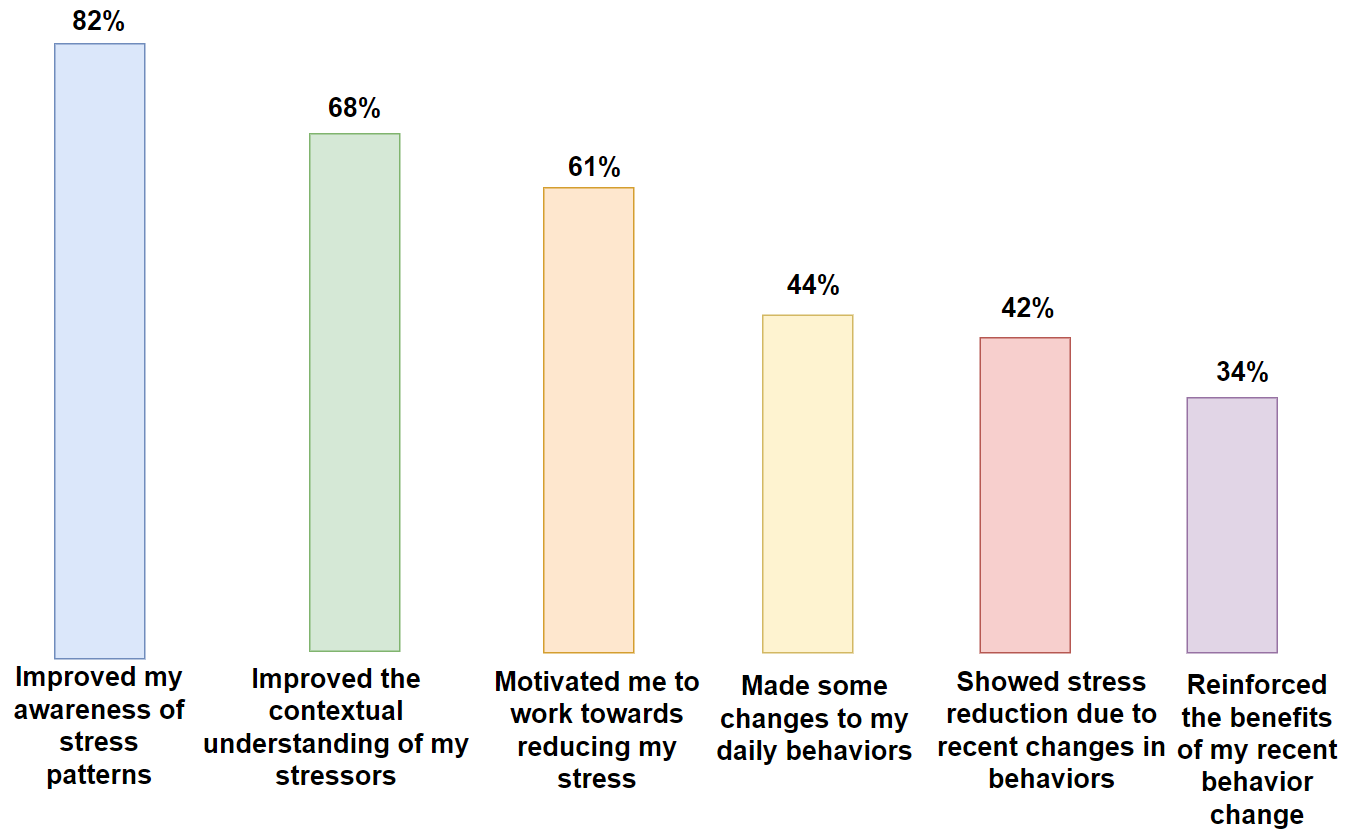}
         \captionsetup{justification=centering}
         \caption{\edit{Weekly Survey Response During MOODS Study}}
         \label{fig:weekly_survey_answer}
     \end{subfigure}
\hfill

        \caption{\edit{Pre-Study and Weekly Survey Response}}
        \label{fig:weekly_stress_ratings}
        \Description{First figure shows the distribution of participants based on their responses for motivation to join the MOODS study. Three responses received over 70\% mentions. Two of these were related to improving
self-awareness and the third was to help advance science.

   The second figure shows the percentage
of participants who reported the specific response (impact of recent visualizations) in one or more of their weekly surveys. A large majority
of participants (82\%) selected either the Improved my awareness of stress patterns or the Improved the contextual understanding of my stressors option. e observe that 61\% of participants mentioned being motivated to work towards reducing their stress. Next, we observe that 44\% of participants mentioned taking specific action to reduce their stress. Finally, 42\% and 34\% of participants respectively mentioned that the visualization helped them see the reduction in stress due to their recent behavior change and reinforced the benefit of their recent behavior change (both referred to as Reinforcement).
}
\end{figure}

\vspace{\baselineskip}

\subsection{Potential Explanations for Overall Reduction in Stress}

\subsubsection{Changes in Self-Awareness of Stress and Stressors During the MOODS Study} \label{sec:self-awareness}
In a pre-study survey, the participants were asked: \texttt{What motivated you to join the \anonymize{MOODS}?}, and to select one or more from five options. Figure~\ref{fig:motivation} shows the distribution of their responses. Three responses received over 70\% mentions. Two of these were related to improving self-awareness and the third was to help advance science.

At the end of each week, the participants were asked: \texttt{What impact did the most recent visualizations have on you?}, and to select all that apply from the options shown on the Figure~\ref{fig:weekly_survey_answer}, and \texttt{None}. A large majority of participants (82\%) selected either the \texttt{Improved my awareness of stress patterns} or the \texttt{Improved the contextual understanding of my stressors} option. Among participants interested in tracking stress patterns (from their pre-study survey), 85\% reported improved awareness, while for those aiming to understand stressors, 74\% reported enhanced contextual understanding of their stressors. 

\edit{This was also supported by comments from participants in exit interviews. $P8$ said \emph{"Eye opener. I was aware before study that often stressed but now after seeing stressors and patterns, it was insightful."} $P18$ said \emph{"I was kinda shocked to see the amount of stressors I had."} $P4$ said \emph{"The study made me more conscious of what stresses me out and how often I feel stressed."} $P10$ mentioned \emph{"It helped me to see how anxious I was getting over minute details."}
$P16$ said \emph{"I was more stressed in the Mornings during the week and big peak around Monday and tapered off throughout the week."}  $P41$ mentioned, \emph{"Increased mental awareness to what repetitively caused my stress."}}

\subsubsection{\edit{Motivation, Action, and Reinforcement towards Self-initiated Behavior Change}} \label{sec:stages-of-change}

\edit{We first quantitatively analyzed self-report data to see how many participants reported motivation, taking action, and receiving reinforcement toward behavior change.} In the weekly survey, participants reported the impact the recent visualizations had on them by selecting a response from the options shown on \edit{x-axis labels} of Figure~\ref{fig:weekly_survey_answer}, and \texttt{None}. 
\edit{Figure~\ref{fig:weekly_survey_answer}} shows the percentage of participants who reported \edit{the specific response in one or more of their weekly surveys.} We observe that 61\% of participants mentioned being motivated to work towards reducing their stress. Next, we observe that 44\% of participants mentioned taking specific action to reduce their stress. Finally, 42\% and 34\% of participants respectively mentioned that the visualization helped them see the reduction in stress due to their recent behavior change and reinforced the benefit of their recent behavior change \edit{(both referred to as \emph{Reinforcement}).}

We next analyze the weekly trend in changes in the proportion of participants who \edit{reported taking an \emph{Action} and receiving \emph{Reinforcement}} (Figure~\ref{fig:beh_trend}). We find a significantly increasing trend (via the MK test) for  \edit{\emph{Reinforcement}} ($b = 13.67$, $m = 0.99\oneS$, $p < 0.05$) and a non-significant, but increasing trend for \emph{Action} ($b = 16.79$, $m = 0.63$,  $p = 0.22$ ). For participants who completed the exit survey, a significant increasing trend was found for both the \emph{Action} ($b = 13.71$, $m = 0.64$, $p =0.08$) and \edit{\emph{Reinforcement}} ($b = 12.79$, $m = 1.22\oneS$,  $p < 0.05$). 

\edit{We next present the four different ways that stressor logging and weekly visualizations motivated participants to move toward or reinforce behavior changes with the corresponding quotes. 

\noindent {\bf 1. Goal Identification:} Some participants identified what to work on. $P28$ said \emph{"I found the simple visualizations that highlighted the data around the frequency of stressors, day of the week, and location to be great in identifying patterns and areas of need."}  $P7$ said, \emph{"The weekly visualizations I received towards the beginning of the study were the ones I liked the most because it was very interesting to see my stress patterns on visual graphs for the first time. Toward the end, they helped me keep track of what factors were causing the most stress so I could focus on reducing them."} $P9$ said, \emph{"I started writing down things to track stressors."} $P42$ highlighted, \emph{"While I was fairly certain that I had stressors occurring in my daily life, this app helped me to acknowledge and attempt the management of those stressors. It also revealed other stress factors that I had not considered before which also informed me of how they could be contributing to a cumulative feeling of stress."}
\begin{figure}[t]
     \centering
     \begin{subfigure}[b]{0.49\textwidth}
         \centering
         \includegraphics[width=
\textwidth]{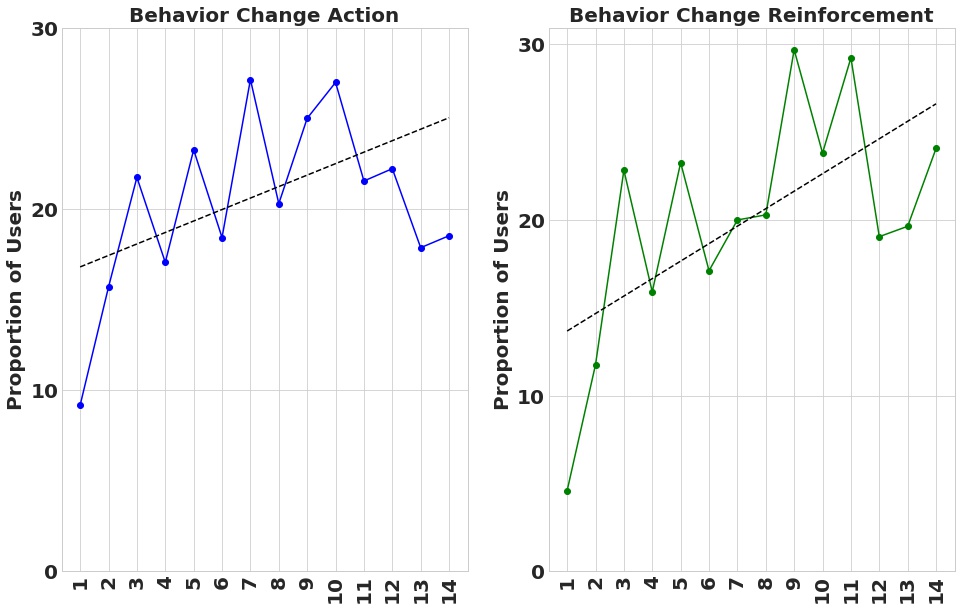}
    \caption{Weekly trend in the proportion of participants being in behavior change \edit{Action and Reinforcement} }
    \label{fig:beh_trend}
     \end{subfigure}
     \hfill
      \begin{subfigure}[b]{0.4\textwidth}
         \centering
             \includegraphics[width=\textwidth]{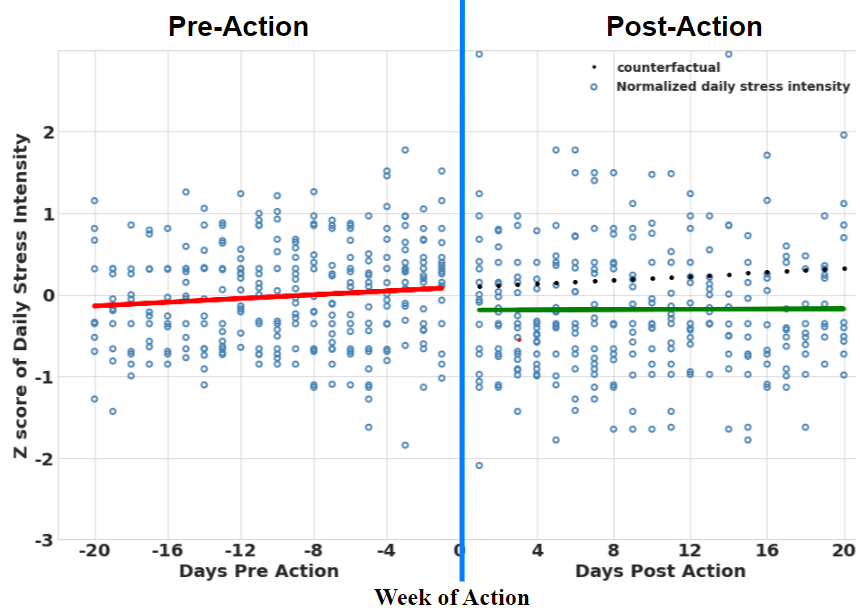}
    \caption{Interrupted Time Series Analysis to assess change in stress levels before \edit{(red line) and after (green line) self-initiated behavior change. }}
    \label{fig:its}
     \end{subfigure}
     \hfill

        \caption{Weekly Trend of Behavior Stages and Comparison of Daily \emph{Stress Intensity} Before and After Initial Action }
        \label{fig:action_yes_no}
        \Description{The first figure shows the weekly trend in changes in the proportion of participants who reported reported taking an Action
and receiving Reinforcement. We find a significantly increasing trend (via the MK test) for Reinforcement and a non-significant, but increasing trend for the Action.

 The second figure shows the interrupted time series analysis to analyze the impact of Self-initiated Behavior Change on Self-reported Stress. We noticed that the majority of the decrease is concentrated in the pre- to post-action time period, i.e., slope was increasing in pre-action stage and the slope is close to zero in post-action stages, with a statistically significant reduction from pre- to post-action. .

}
\end{figure}

\noindent {\bf 2. Identify Dominant Stressor(s):} Some participants learned about their dominant stressors. $P26$ said \emph{"I realized my biggest stressor was approaching the deadline and tried not to procrastinate as much and work on it."}  $P8$ said \emph{"I noticed patterns across the time of the day. Example: trouble sleeping (stressor)."} $P4$ mentioned \emph{"Although it was hard to confront as I realized that most of my stress is caused by other people (interpersonal conflict), which is pretty annoying."} $P25$ said \emph{"Change or communicate with my advisor about being responsible for most of my stress."}

\noindent {\bf 3. Identify Most Stressful Contexts:} Some participants learned about their most stressful contexts. $P27$ said, \emph{"I realized I was most stressed while working and became more aware of certain situations where more stressed especially the meetings."} $P31$ said \emph{"I identified some stressors that I can work on to reduce, especially social contexts."} 

\noindent {\bf 4. Reinforcement of the Behavioral Change Benefits:} The visualization helped some participants reinforce their behavioral changes by showing gradual stress reduction. $P7$ said \emph{"I noticed a decrease in my stress on the road and behind the wheel!"}. $P8$ said, \emph{"I noticed less stressed episodes towards the end compared to the beginning of the study."} $P10$ said \emph{"I loved the weekly trends because I could see if what I have been doing has helped in the long run."}
}

\subsubsection{\edit{Types of Self-Initiated Behavior Changes}} \label{sec:behavioral-change-types}
\edit{During the exit interview, we asked participants to describe any behavioral changes they made during the study. They reported making fourteen (14) behavioral changes that we broadly classify into three groups --- 1.) Better regulating emotions before, during, or after stressful situations, 2.) Improving productivity via enhancing focus and better planning, and 3.) Taking better self-care via developing healthier habits. These are described in Table~\ref{tab:behavior_changes_types}, along with supporting quotes from participants.}

\begin{table}[t]
\caption{\edit{Fourteen (14) Self-Initiated Behavior Changes Reported by Participants in Exit Interview}}
    \centering
    \resizebox{\textwidth}{!}
      {
      {\renewcommand{\arraystretch}{2}%
\begin{tabular}{|c|c|c|}
    \hline
\edit{\textbf{\huge Groups} }  &\edit{\textbf{\huge Behavior Changes (14)}}  & \edit{\textbf{ \huge Supporting Quotes from Participants }}   \\
    \hline

            & \edit{ \makecell[l]{\Large Staying calm} }        &   \edit{\textit{\makecell[l]{\Large P43 : "I tried to stay calm during situations or  try to meditate through them." }} }          \\  \cline{2-3}
            \edit{\textbf{\makecell[l]{\Large Emotion Regulation}}  }       &\edit{\makecell[l]{\Large Becoming less reactive by \\ \Large controlling  their
emotions} }   & \edit{\textit{\makecell[l]{\Large P12 : "I am now controlling emotions  in certain situations which I never thought earlier."}   }}           \\  \cline{2-3}
             & \edit{ \makecell[l]{\Large Anticipating stressful situations}  }       &    \edit{\textit{\makecell[l]{\Large P12 : "I anticipate and think about the stress levels in advance while going through   some situations."}}  }               \\  \cline{2-3}
            & \edit{ \makecell[l]{\Large Breaking away from stressful \\ \Large situations} }        &   \edit{\textit{\makecell[l]{\Large P42 : "Most of my self-improvement goals were to find moments to breakaway  from  stressful situations  when they  \\ \Large  happen or to pre-plan physical  \Large activities   that would be after a known  stressful situation."}} }               \\  \cline{1-3}

            & \edit{\makecell[l]{\Large Slowing Down} }   &    \edit{ \textit{\makecell[l]{\Large P8 : "I used to rush in work but now in the last month, I have started to slow down."}} }               \\  \cline{2-3}
                 \edit{ \textbf{   \makecell[l]{\Large Enhanced focus, \\ \Large Better planning, and\\ \Large   Improved productivity}}}         & \edit{\makecell[l]{\Large Better planning}    }      &    \edit{\textit{\makecell[l]{\Large P26 : "I started taking more time to plan the day more so as to reduce stress." \\ \Large P35 : "I made changes to my schedules, particularly on Thursdays, to  improve my daily organization and planning."  }}    }              \\ \cline{2-3}
        &  \edit{ \makecell[l]{\Large  Goal setting } }    &   \edit{ \textit{\makecell[l]{\Large P13 : "I set myself a goal to be more consistent with my work.  I could have done better but  I did improve a little." }} }    \\  \cline{1-3}

       & \edit{ \makecell[l]{\Large Eating regulation}}      &         \edit{    \textit{\makecell[l]{\Large P9 : "I started
writing down things to track stressors, installed  an app to document eating." }}  }      \\  \cline{2-3}
        
          &  \edit{ \makecell[l]{\Large Taking more walks  } }      &    \edit{  \textit{\makecell[l]{\Large P4 : "I started working out more and taking walks to increase my step count." }}    }            \\ \cline{2-3}
          \edit{ \textbf{ \makecell[l]{\Large Better Self-care and \\ \Large Making Healthier \\ \Large Choices \\ {}} }}    &     \edit{\makecell[l]{\Large Improving sleep} }      &    \edit{ \textit{\makecell[l]{\Large P30 : "I tried to lower the
amount of stress for myself at home. I got more sleep and made caring for myself a priority." \\ \Large P10 : "My goal was to make sure I did not stress as much with smaller details. I did my best to get adequate sleep , \\ \Large which lowered my stress."
            }}}               \\  \cline{2-3}
             &   \edit{\makecell[l]{\Large Taking more breaks between \\ \Large work  }  }    &      \edit{\textit{\makecell[l]{\Large P28 and P8 : "I started taking more breaks between
work through the day." }}     }           \\\cline{2-3}
              &  \edit{  \makecell[l]{\Large Deep breathing }  }    &    \edit{  \textit{\makecell[l]{\Large P28 and P13 : "I started practicing deep breathing more frequently throughout the day."}}}                \\\cline{2-3}
               &   \edit{ \makecell[l] {\Large Meditation } }    &      \edit{\textit{\makecell[l]{\Large P42 : "Most
of my behavioral changes centered around taking more walks or providing time to stop what  I was doing  \\ \Large and   to
breathe  or meditate  to settle out of a stressful situation." }}  }              \\\cline{2-3}
                &  \edit{  \makecell[l]{\Large Quit smoking}  }     &    \edit{  \textit{\makecell[l]{\Large P23 : "I quit smoking."}}}                \\
    \hline
\end{tabular}}}
\label{tab:behavior_changes_types}
\Description{The Table shows 14 different behavior changes self-initiated by participants as reported in the exit interview along with the supporting quotes from participants. We group them into 3 groups: Emotion Regulation, Enhanced focus, Better planning and Improved productivity, and Better self-care and Making healthier choices. Emotion Regulation includes behavior changes such as staying calm, becoming less reactive by controlling emotions, anticipating stressful situations and breaking away from stressful situations.  Enhanced focus, Better planning and Improved productivity includes slowing down, better planning and goal setting. Better Self-care and Making healthier choices includes behavior changes such as improving sleep, eating regulation, taking more walks, taking more breaks between work, deep breathing, meditation and quit smoking. Some examples of supporting quotes include P4 said "I started working out more and taking walks to increase my step count."and  P12 said "I anticipate and think about the stress levels in advance while going through some situations."}
\end{table}

\subsubsection{Impact of Self-initiated Behavior Change on Self-reported Stress} \label{sec:interrupted-time-series}
As previously noted, approximately 44\% of participants in the MOODS study reported making self-initiated changes to reduce their stress. We subsequently focused on examining whether these actions had a noticeable effect on reducing their daily stress levels. To do this, we selected those participants who initiated such actions for the first time between the fourth and eleventh week of the study and participants who had stress intensity ratings for at least one week before and one week after adopting these stress-reducing actions. This process resulted in a subset of 17 participants meeting these criteria. On average, these participants began their stress-reduction actions in the sixth week, with a standard deviation of 2 weeks.

An MK test on the weekly \emph{Stress Intensity} data of these participants yields $b=1.81$ and $m=-0.03$, where the intercept is slightly higher and the slope slightly lower than the population. To assess the effect of their self-initiated change on their stress ratings, we treated the first week they mentioned taking an action as the week of their intervention \edit{(action week)}. We organized the time series data of each of these participants so that their pre-action and post-action weeks are aligned. To account for variations in stress intensity among participants, we normalized individual stress intensity scores by converting them into $z$-scores for the purpose of this analysis. \edit{We calculated a z-score for each individual's overall study stress intensity.} We excluded the action week's data from this analysis. Figure~\ref{fig:its} shows the \emph{interrupted time series} analysis, \edit{a quasi-experimental design to analyze data over an extended timeframe before and after an intervention to assess the impact of the intervention~\cite{kontopantelis2015regression} in natural experiments and studies without randomization. These effects can manifest as alterations in the time series trend or shifts in the intercept following the intervention. The $y$-axis of Figure~\ref{fig:its} depicts normalized daily stress intensity scores. These scores are depicted from the 20\textsuperscript{th} annotation day before the action week on the left side of the blue line. Additionally, the normalized stress intensity scores from after the action week up to the 20\textsuperscript{th} annotation day are shown on the right side of the blue line. Dotted lines represent counterfactual (stress levels had the self-initiated behavior change not happened.) Although the MK test showed a decreasing trend, we notice that the majority of the decrease is concentrated in the pre- to post-action time period, i.e., slope was increasing in pre-action stage and the slope is close to zero in post-action stages, with a statistically significant reduction ($-0.278\oneS, p <0.05$) from pre- to post-action. }

\begin{figure}[t]
     \centering
     \begin{subfigure}[b]{0.31\textwidth}
         \centering
         \includegraphics[width=\textwidth]{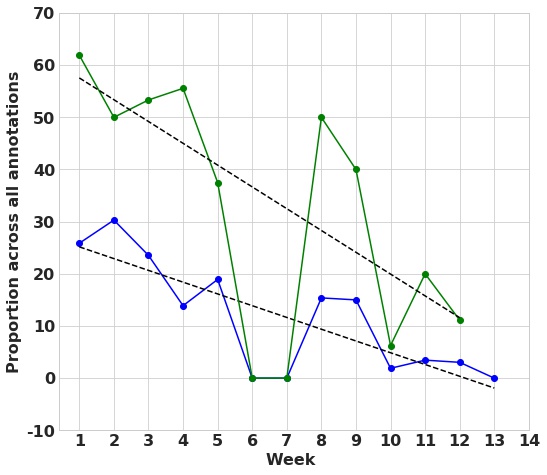}
         
         \captionsetup{justification=centering}
         \caption{Proportion of transportation-related stressors (blue) and stressors reported during commute (green) for $P38$}         \label{fig:commute}
     \end{subfigure}
     \hfill
      \begin{subfigure}[b]{0.33\textwidth}
         \centering
         \includegraphics[width=\textwidth]{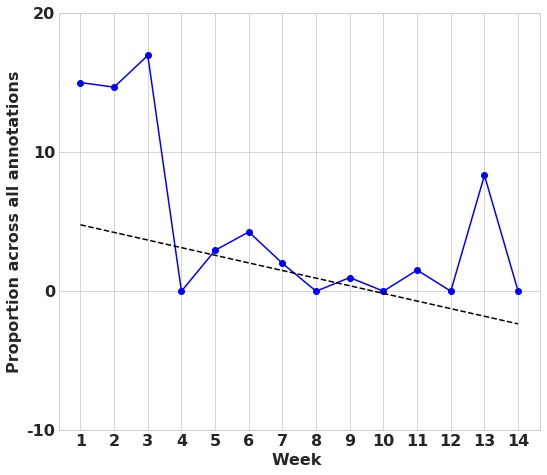}
         \captionsetup{justification=centering}
         \caption{Proportion of work-related stressors for $P8$}
         \label{fig:work_issues}
     \end{subfigure}
     \hfill
     \begin{subfigure}[b]{0.35\textwidth}
         \centering
         \includegraphics[width=\textwidth]{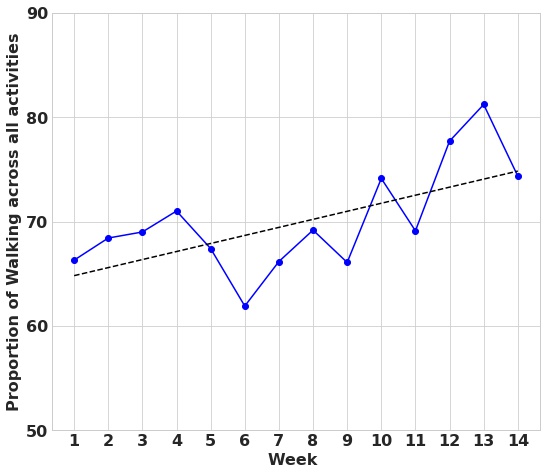}
         \captionsetup{justification=centering}
         \caption{Change in the amount of walking activity for $P42$}
         \label{fig:walking}
     \end{subfigure}
     
        \caption{Data supporting evidence of specific behavioral changes reported by three participants during exit interview}
        \label{fig:behavior}
        \Description{The first figure shows the weekly proportion of transportation-related stressors reported by P38 during their commute. The figure indicates that as the weeks progressed, this participant reported experiencing fewer transportation-related stressors.
The second figure shows the weekly trend of work issues and stressors across all annotations for P8. The figure shows a significantly decreasing trend in the frequency of work-related stressors over time.
The third figure shows the weekly walking trend of P42 who stated that their primary behavioral changes revolved around increasing their walking activity. As depicted in the figure, there is a noticeable and significantly upward trend in their walking activity. 
}
\end{figure}

\subsection{Baseline Stress Differences Between Participants with and without Self-Improvement Goals or Actions} \label{sec:baseline-stress}

During the exit interview, we asked participants if they had set any self-improvement goals. Out of 46 participants who completed the exit interview, 18 reported having set self-improvement goals either before or during the MOODS study. We also asked participants if they had implemented any behavioral changes over the course of the study. Out of 46 participants, 24 acknowledged making behavioral changes as a result of their study involvement. Among the cohort of participants who had self-improvement goals before the study commenced, 88.8\% actively engaged in making behavioral changes throughout the study duration. In contrast, out of the participants who did not report any goals prior to the study, only 29\% reported making behavioral changes during the study period.

We wanted to explore whether participants who had established goals or made behavior changes reported higher or lower stress levels at the outset compared to participants without any goals or those who did not report making any behavioral changes during the study. Since baseline stress measurements were not initially available, we estimated a baseline period for analysis. This baseline period was determined by identifying the week at which participants reported significantly different stress levels in comparison to the initial week(s).
By employing the Wilcoxon signed rank test, we conducted a statistical comparison between the stress intensity levels of the first week and those of subsequent weeks. We find a significant reduction in stress intensity by the fourth week (with mean values of 1.81 in the first week and 1.57 in the fourth week, $p < 0.05$). Consequently, we identified the baseline period as spanning the first three weeks. Thus, we define baseline stress intensity for each participant as the average stress intensity in the first 3 weeks. We employed the Mann-Whitney test to comparatively assess baseline stress intensity among different participant groups based on whether they had established goals prior to the study or had pursued behavioral changes during the study.

\subsubsection{Do participants who reported having specific self-improvement goals report a higher baseline stress?}
We examined the average baseline stress intensity among participants who had set goals before participating ($n=18$) and those who had not ($n=28$). The results indicated that participants who had established goals prior to the MOODS study exhibited an average baseline stress intensity of $1.83$ vs.\  $1.67$ for those who did not set any goals. Although not statistically significant ($p=0.32$), participants with pre-established goals tended to report a slightly higher baseline stress intensity. 

\subsubsection{Do participants who reported making behavioral changes during the MOODS study report a higher baseline stress?}
In the examination of baseline stress intensity among participants, it was found that individuals who made behavioral changes during the study ($n=24$) exhibited an average baseline stress intensity of $1.89$ vs.\ $1.56$ for those who did not make any behavioral changes. Although not statistically significant ($p=0.15$), participants who engaged in behavioral changes reported a slightly higher baseline stress intensity.

\subsection{Illustrations of Data Supporting Behavioral Changes Reported in the Exit Survey} \label{sec:behavioral-changes-data}

During the exit interview, we asked participants for more details if they mentioned having a goal or making behavior changes. \edit{We analyze how well these changes are reflected in the data of some participants for illustration purposes.}

$P38$, \edit{whose most prevalent stressor was transportation-related,} mentioned \emph{"I had a goal of being less stressed during their commute by utilizing deep breathing and being cognizant of my listening choices on the drive."} Figure~\ref{fig:commute} shows the weekly proportion \edit{(percentage)} of transportation-related stressors reported by this participant. The blue line represents the proportion across all annotations related to transportation stressors, while the green line specifically represents the proportion of annotations made during their commute. The commute data was extracted from the Google Activity API. \edit{On Weeks 6 and 7,  no transportation-related stressors were reported (Figure~\ref{fig:commute}).} MK test shows a significantly decreasing trend in the weekly proportion of transportation-related stressors across all annotations ($b= 27.42 $, $m= -2.25\oneS$, $p < 0.05$) and for stressors reported during the commute ($b= 61.73 $, $m= -4.47\oneS$, $p < 0.05$). This indicates that as the weeks progressed, this participant reported experiencing fewer transportation-related stressors.

$P8$, \edit{whose most prevalent stressor was work issues,} stated \emph{"I tried to work a little less and take more breaks during work, and as a result, noticed fewer episodes of stress towards the end of the study compared to the beginning."} Figure~\ref{fig:work_issues} shows the weekly \edit{proportion of} \texttt{work issues} stressors across all annotations. The MK test shows a significantly decreasing trend ($b= 5.32 $, $m= -0.54\oneS$, $p < 0.05$) in the frequency of \texttt{work-related} stressors over time. 

Figure~\ref{fig:walking} shows the weekly walking trend of $P42$  who stated \emph{"My primary behavioral changes revolved around increasing my walking activity."} As clearly depicted in the figure, there is a noticeable and significantly upward trend ($b= 64.05 $, $m= 0.77\oneS$, $p < 0.05$) in their walking activity which was also extracted from the Google Activity API.

\subsection{Burden \& Utility   In Using The App}
\begin{figure}[t]
    \centering
    \includegraphics[width=0.99\textwidth]{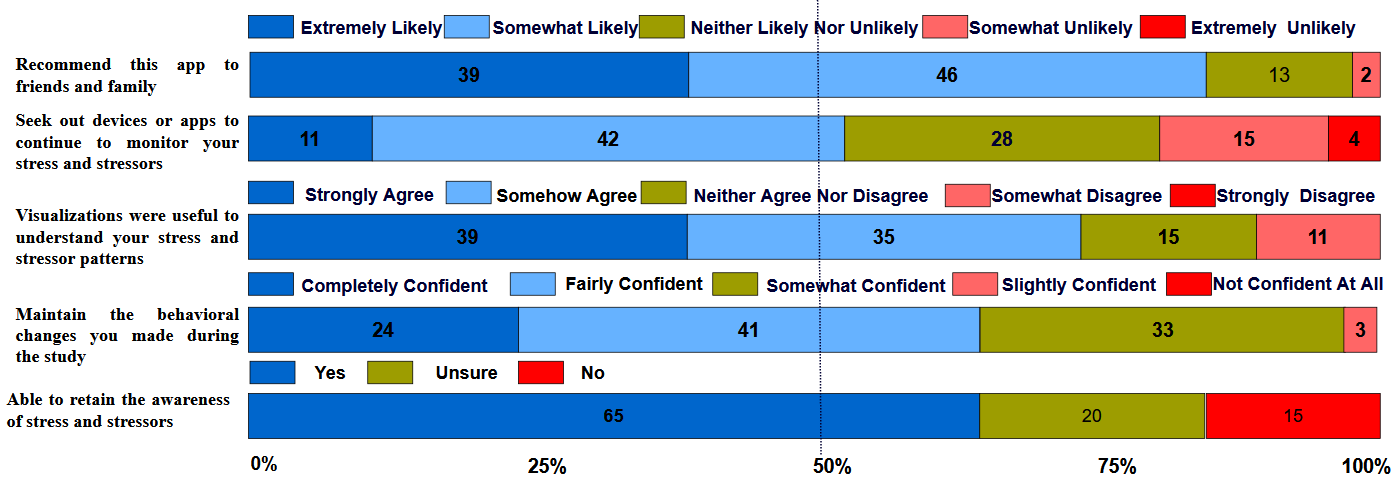}
    \caption{Participants' Evaluation and Perception of the Stressor Logging Application }
    \label{fig:exit_survey}
    \Description{   
    To better understand the utility factor behind high participant engagement in this study, we analyze their responses to five questions from the exit survey. The figure shows that visualizations helped the majority (about 75\%) of the participants to understand stress and stressor patterns while 65\% said they would be able to retain awareness of stress and stressors. Moreover, 100\% of the respondents were more or less confident about maintaining the behavioral changes they initiated to manage their stress which might be the reason that a little over 50\% are likely to actively seek out such stress tracking apps in future. Overall, the majority of the participants found the app to be useful as more than 75\% said they would recommend this to friends and family which is a good indicator of the perceived utility potentially driving the high participant engagement.}
\end{figure}

\subsubsection{Participant Burden} \label{sec:burden}
When asked in the exit interview, approximately 84\% and 78\% of the participants stated that the MOODS study app and smartwatch did not disrupt their daily activities respectively. We next \edit{analyze the physical and mental effort involved in using} the app \edit{i.e., burden}.

\begin{figure}[t]

     \hfill
     \begin{subfigure}[b]{0.45\textwidth}
         \centering
         \includegraphics[width=\textwidth]{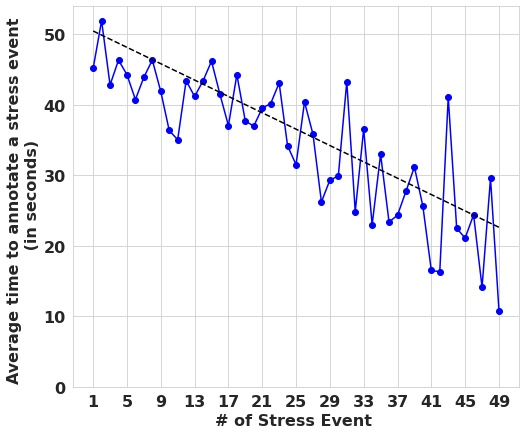}
         
         \captionsetup{justification=centering}
         \caption{Time taken to annotate a stressor}         \label{fig:avg_time}
     \end{subfigure}
     \hfill
     \begin{subfigure}[b]{0.43\textwidth}
         \centering
         \includegraphics[width=\textwidth]{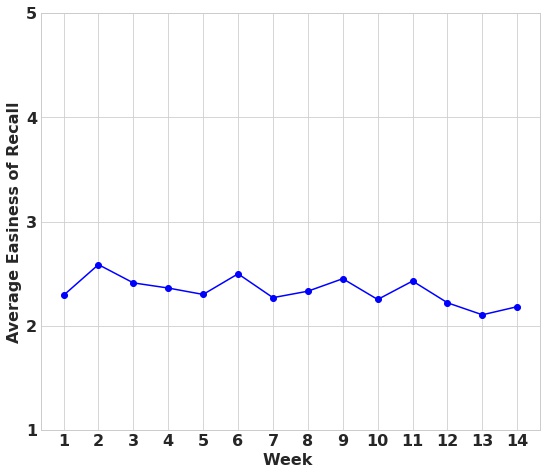}
         \captionsetup{justification=centering}
         \caption{Ease of Recalling Stressors}
         \label{fig:burden}
     \end{subfigure}
     
        \caption{Time commitment, and recall burden}
        \label{fig:avg_time_burden}
        \Description{
The first figure shows the average time it took participants to annotate their stress episodes. 
By implementing a predictive text input module that presented potential stressors as participants typed, there was a significant reduction in the time taken for stressor entry as more episodes were annotated.
The second figure shows the weekly score of ease of recollecting their stressors. The score remains closer to 2, indicating that recalling the stressors did not pose a burden, perhaps due to asking them right after an event occurred and showing them the time and location.

}
\end{figure}

Aside from charging the smartwatch, participants were asked to rate their stress and annotate the stressor multiple times daily. Participants reported in~\cite{lee2020toward} that typing an event repeatedly without auto-complete became burdensome. Upon their recommendation, we implemented a predictive text input module that presented potential stressors as participants typed. \edit{We used logs collected from the smartphone app to calculate the time taken to enter a stressor for every stressor entry. The predictive text input module} resulted in a significant reduction in the time taken for stressor entry over time as more and more episodes were annotated ($b = 50.46 $, $m = -0.58\oneS$, $p < 0.05$) (see Figure~\ref{fig:avg_time}).  The gradual reduction in time commitment may have played a role in keeping the participants engaged for a long duration. 

Another aspect of the burden may be the mental effort in recalling the stressors. In the weekly survey, participants were queried about the ease of recollecting their daily stressors from the past week, and their responses were quantified on a numerical scale (ranging from "Quite easy" to "Very difficult," scored from 1 to 5). As Figure~\ref{fig:burden} shows, the score remains closer to 2, indicating that recalling the stressors did not pose a burden, perhaps due to asking them right after an event occurred and showing them the time and location.

\subsubsection{Participant Ratings of App Utility} \label{sec:feedback}

We analyze the responses to five questions about the app's utility from the exit survey. Figure~\ref{fig:exit_survey} shows the five questions and the proportion of participants selecting specific responses. The scores indicate that participants were very satisfied with the app and reported deriving tangible real-life benefits by visualizing their own data. In fact, 85\% participants indicated a willingness to recommend the app to friends and family.

Comments from some participants explain their sentiments. $P38$ said \emph{"This has been an interesting experience. I encouraged others to participate as well."} $P28$ said \emph{"I really enjoyed participating in this study as I know it likely provides valuable information; however, it was personally rewarding as well to help me identify my stressors and areas of need.  So from that personal standpoint I appreciate the opportunity to participate."}

\vspace{-0.3\baselineskip}

\section{Discussion \& Limitations}

\subsection{Feasibility of Logging Stressors}
\edit{As stressor logging involves both physical and mental effort, engaging participants so they keep logging over a long time is a challenge.} An analysis of mental health apps (with 10,000 or more installs) found that the average retention rate at the 30-day mark was 3.3\%\cite{baumel2019objective}. The most frequent reasons for discontinuance include too much time needed to enter data, loss of interest, hidden costs, and confusing to use~\cite{krebs2015health}. \edit{Although our study did not involve any costs, it could have lost participants due to the other three factors. However, our study showed that using novel mechanisms to lower burden and improve utility resulted in a 25 times higher retention rate (see Section~\ref{sec:user-retention-contribution}). We also found that 85\% of participants were willing to recommend our apps (see Figure~\ref{fig:exit_survey}). We hypothesize that this high retention rate and high approval rating resulted from the following --- 1) reducing the mental effort of recalling stressors by prompting them close to the time of occurrence of a potentially stressful event, but waiting until the event was over so they can recover; 2) reducing the physical effort in entering the stressor by using predictive auto-complete. As fatigue grows over time, predictive auto-complete reduces the time needed for stressor logging as most stressors reported are already in the database; 3) integrating stressors with stress arousal and spatiotemporal context in visualizations so they can identify their dominant stressors and contexts; and 4) providing new types of visualizations each week that allowed them to progressively dig deeper in their data.} 
\vspace{-0.5\baselineskip}

\subsection{Impact of Stressor Logging and \edit{Integration of Stressors in Self-reflective Visualizations }}\label{sec:behavioral-changes}

\edit{To enhance stress management, previous studies have employed various new technologies, including chatbot-triggered exercises rooted in Cognitive Behavioral Therapy (CBT) and Dialectical Behavioral Therapy (DBT)~\cite{howe2022design}, a browser-based application delivering micro-stress interventions~\cite{tong2023just}, a calendar-mediated stressor anticipation application~\cite{lee2020toward}, and a system focused on setting adaptive goals in Exercise, Meditation, and Accessibility dimensions~\cite{konrad2015finding}. 
Our study shows that sensor-prompted logging of stressors and integrating stressors with associated spatiotemporal contexts in stress visualizations for self-reflection holds a potential for a statistically significant decrease in self-reported stress that is at par or better than many stress intervention studies. We note that people report having \emph{some} or \emph{little control} over their daily stressors~\cite{cerino2023perceived}, implying that there is only a small subset of stressors that can be addressed via behavioral modifications. Therefore, a reduction of 10\% in self-reported stress intensity and 10 fewer stressors per month reflects a meaningful change that can potentially improve long-term health outcomes~\cite{almeida2002daily,ullrich2002journaling}.

Exploration of the mechanisms behind the stress reduction shows that stressor logging and weekly visualizations helped participants become more aware of patterns in their stress and stressors, identify what they needed to work on (e.g., dominant stressors or most stressful spatial or temporal contexts), and evaluate the impact of self-initiated behavioral changes to discover what worked best for them. Specifically, participants reported making fourteen (14) types of behavioral changes (see Section~\ref{sec:behavioral-change-types}). It provides support for offering a large repository of interventions~\cite{tong2023just} and provides support for anticipatory interventions~\cite{lee2020toward} in addition to during and post-stress interventions, as participants reported using each type. Finally, our study shows that if stress intervention can indeed match the source of stress as participants asked for in~\cite{howe2022design}, it has a chance to produce a longer-lasting reduction in stress.}

\subsection{Design Implications for Stress Tracking Apps}

\subsubsection{Focus on Stressors} 
One of the findings from this study is that providing an easy and quick way to log stressors right around the time a stress event may be occurring can improve the chances of engaging participants in logging their stressors. Participants mentioned the approach followed in this study to log their stressors easy to recall (see Figure~\ref{fig:avg_time_burden}).

\subsubsection{Insightful Visual Summaries}
Once participants log their stressors, helping them visualize their own data improved the awareness of stress and stressor for most participants. It led many of them to make behavioral changes, improve productivity, and adopt healthier habits. This study also showed which kinds of visualizations of stressors were found to be most and least helpful by the participants. Adopting these can improve the real-life utility of stress-tracking apps.

\subsubsection{Matching Interventions to Sources of Stress}
It is desirable to find stress interventions that can work for many people in a variety of situations. However, a recent study~\cite{howe2022design} found that participants wanted the intervention to relate to the specific stressor they were experiencing. Our findings support this as well. Even though our study did not recommend any intervention, engagement with stressor logging and visualization and the resulting growth in awareness led many participants to initiate several behavioral changes. When we analyze the specific behavioral changes initiated (see Section~\ref{sec:behavioral-changes}), we find substantial diversity. These show that stress interventions that include a diverse array of suggestions~\cite{tong2023just} and aim to match the intervention to the source of stress are more likely to be effective. \edit{For an intervention application to automatically match a prompted intervention to the source of stress, however, AI models are needed that can identify the stressor from sensor data. Data collected in our study that have both field-collected sensor data and stressor labels provided by participants can accelerate such research.}

\subsection{Limitations}
This study had several limitations that can inspire future research.

\edit{
\subsubsection{Benefit of Using Sensor-triggered Prompts}
In this study, prompts were generated across all spectrum of stress scores so sensor data could be collected to further improve the performance of AI models for detecting stress. Future studies could reduce participant burden further by only selecting higher stress arousal scores to generate prompts. To establish how useful it is to use such pre-trained AI models to generate prompts, future studies can conduct a randomized control trial (RCT), wherein one group receives prompts tailored for higher stress scores to minimize participant burden, while the other group uses an alternative mechanism to reduce participant burden, i.e., receive randomly generated prompts without having to wear any smartwatch. The objective would be to assess and compare the effectiveness of these two approaches in terms of user engagement, retention rates, and stress reduction.

}
\subsubsection{Observational Study with No Control Arm} 
We observed an approximately 11\% reduction in self-reported stress over the 14-week study period with momentary stressor logging and weekly visualization of self-reported stressors. This study-long reduction in self-reported stress is higher than several recent works on stress intervention. However, this should only be regarded as showing a potential for stress reduction and not definitive evidence due to several caveats. First, this was an observational study with participants' own data from the initial weeks acting as their control. Second, this being a study without any ongoing or end-of-study incentive, it may have attracted participants motivated in stress tracking. Third, there was no micro-incentive for completing self-ratings or weekly surveys. This may increase the chance that missing data may not be random. Fourth, different participants contributed data for different amounts of time potentially introducing other biases. To see if the trends persist when the analysis was limited to $4\leq\forall k\leq 13$ weeks resulting in more participants (up to $n=86$) available for complete case analysis, we repeated the LMM analysis and found similar trends with $-0.023\leq m\leq -0.029$ for \emph{stress intensity} and $-0.018\leq m\leq -0.029$ for \emph{stress frequency}. Finally, the magnitude of reduction in stress was estimated from a linear fit and it was not using a standard scale such as the Perceived Stress Scale (PSS) at both pre- and post-study stages. A randomized control trial (RCT) is needed to truly determine whether momentary stressor logging and weekly visualizations can indeed reduce stress and improve awareness leading to behavior change and should become an essential component in future stress interventions.

\edit{
\subsubsection{Disentanglement of Visualizations to Isolate their Impact}

In this study, several factors could have potentially contributed to the reduction in self-reported stress intensity and frequency. They included getting prompts upon the conclusion of a potentially stressful event, logging stressors, having access to the app where they can review a timeline of their self-reported stressors, getting 16 types of visualizations that included a combination of stress, stressors, spatial context, and temporal context aimed at eight insights, the introduction of a new visualization type each week, and weekly surveys. A future study can investigate the incremental impact of one or more of these factors and identify the ones with a potential for the highest impact so they can be considered for future sensor-triggered interventions.
}

\subsubsection{No Distinction between Productive and Unproductive Stressors} 
Stress is sometimes necessary to stay alert, overcome challenges, and improve work performance. Examples include challenging projects, competitions, and exams. Unproductive stressors harm well-being, causing negative impacts on health and efficiency. Examples include work overload, conflicts, and catastrophization. In this study, no distinction was made to account for this. $P16$'s experience in a study exemplifies this. Engaged in a dissertation, they were annoyed by watch notifications for stress. Recognizing stress from academics, they disabled notifications, seeing stress as intrinsic to their goals. They even removed the watch to avoid interruptions. This underscores understanding of participant context for effective stress management. Future stress management apps should aim to distinguish between productive and unproductive stressors.

\subsubsection{Positive vs.\ Negative Emotions}
There might be many instances where \edit{physiological} responses are triggered by excitement, eating, or substance consumption. Such instances could be interpreted as positive stress or eustress. \edit{Eustress is a positive
cognitive response to a stressor, which is associated with
positive feelings and a healthy physical state 
whereas; distress is defined as the stress associated with negative feelings and 
physical impairments~\cite{lazarus1993psychological}).} While the identification of eustress instances could be enlightening for participants, some participants in this study raised concerns about such instances. $P4$ said \emph{"Any time I so much as watch a TV show with drama or tension, the watch registers a suspected stress episode. Excitement, exercise, and substance use all trigger the app."} $P41$ said \emph{"I personally was confused on how actual stress was being detected vs.\ just an elevated heart rate. The app would alert to stress during times of relaxation. For example, while watching TV, eating, or just general excitement about something."} Such testimonials indicate that prompting participants during eustress moments could confuse them and reduce their trust in the stress app. Developing methods to automatically differentiate eustress from distress  can further improve the utility of stress apps.

\subsubsection{Device Limitations}
As shown in Figure~\ref{fig:exit_survey}, participants were not as enthusiastic about seeking out devices or apps. When asked about disliked aspects of the app, the most frequent complaint was the short battery life of the smartwatch as it lasted about 6 hours. Fortunately, this issue is easily addressable in commercial apps as modern smartwatches last the entire day. Some iPhone participants highlighted occasional Bluetooth connectivity issues. Intermittent problems in establishing and maintaining a stable connection between the smartwatch and iPhone frustrated these participants. 

\subsubsection{Stress Rating Biases}
The likelihood of a presented event considered to be stressful by the smartwatch app was displayed in the smartphone app when participants were asked to annotate the event. It was added at the request of pilot participants so they could more easily remember a stressor. $P31$ said \emph{"It influenced sometimes but many times it made me think harder to recall the stressors."} $P35$ said \emph{"Stress likelihood didn't influence and might have been subconsciously."} Although some participants mentioned that this did not bias them ($P26$ said \emph{"I had no bias from stress likelihood."}, $P27$ said \emph{"It may have had a little bias but I tried to rely on my own judgment."}), this display may have created a bias in the stress ratings participants chose for the event. As our analysis to determine a trend in this data included a within-person analysis, this bias may not have had a substantial impact on the results presented here. 

\subsubsection{No Analysis of Clinical Significance}
In our analysis, we only tested for statistical significance. Establishing any clinical significance (i.e., impact) of momentary stressor logging and visualizations will require pre and post-study assessment of clinical outcomes of interest such as addiction, depression, weight, sleep, daily symptoms, etc.

\section{Conclusion}
Real-time detection of \edit{physiological events} indicative of potential stress in conveniently-worn wearables opens exciting new possibilities to improve stress management. Research presented here in a 100-day study shows the feasibility of soliciting stressors from participants upon the detection of a physiological event by a smartwatch. Our findings show that momentary stressor logging and providing simple and insightful visualizations of these data can lead to better awareness of stress and stressors. It can also lead to self-initiated behavior modifications. Future work can investigate how to integrate such capabilities in consumer wearables and conduct micro-randomized trials to determine how such capabilities may be used in conjunction with momentary interventions to help people better manage their stress. 
\section*{Acknowledgements}
We thank the anonymous reviewers for greatly improving the paper. We also thank Joseph Biggers from the mDOT Center at the University of Memphis for helping with participant recruitment. Research reported here was supported by the National Institutes of Health (NIH) under award P41EB028242. It was also supported by the National Science Foundation (NSF) under awards ACI-1640813 and CNS-1822935. The opinions expressed in this article are the authors' own and do not reflect the views of the NIH or NSF.

\bibliographystyle{ACM-Reference-Format}
\bibliography{sample-bibliography}
\end{document}


\setcopyright{acmcopyright}
\copyrightyear{2024}
\acmYear{2024}
\acmDOI{}

\acmConference[CHI '24]{Conference on Human Factors in Computing Systems}{May 11--16, 2024}{Honolulu, Hawaii}
%
%
\acmBooktitle{CHI '24: Conference on Human Factors in Computing Systems,
 May 11--16, 2024, Honolulu, Hawaii} 

\maketitle



We designed 16 visualizations of stress ratings and stressor logs that participants provided to facilitate their self-reflection. Details of each visualization with an example appear below.

\section{Week 1}
\subsection{Overall Average}

\begin{figure}[h]
     \centering
     \begin{subfigure}[b]{0.45\textwidth}
         \centering
         \includegraphics[width=6cm]{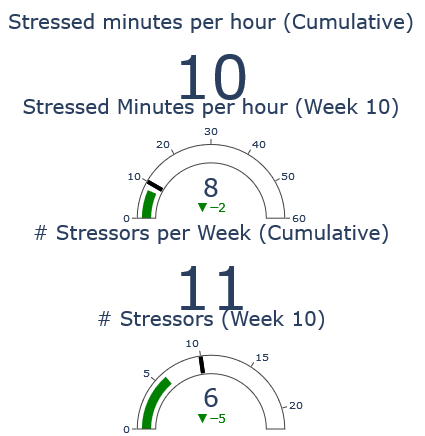}
\caption{Overall and recent trends in stress and stressors }
         \label{fig:overall}
     \end{subfigure}
     \hfill
      \begin{subfigure}[b]{0.45\textwidth}
         \centering
         \includegraphics[width=6 cm]{figs/prominent_stressor_context.png}
\caption{Context (place and time) of five most prominent stressors }
         \label{fig:Prominent}
     \end{subfigure}
     \hfill

        \caption{Visualizations for Week 1}
        \label{fig:vis_1}
        \Description{
}
\end{figure}

Participants received two sets of visualizations at the end of their first week. The first one  (see Figure~\ref{fig:overall}) presented an overall summary of their data. It included
\begin{enumerate}

   \item Stressed Minutes per Hour (aggregated from all data collected thus far): This metric represents the average duration of stress experienced per hour (of data collected). As the amount of data collected can vary between participants and between days for the same participant, per hour instead of per day was used.

   \item Average Stressed Minutes (in the Current Week): To contrast with the overall average, this value indicates the average duration of stress experienced in the most recent week. It offers a snapshot of the ongoing stress levels during the current week.

   \item Average Number of Stressors Reported per Week (aggregated from all data collected thus far): The average number of stressors reported per week thus far indicates the typical frequency of stressors in the participant's daily life.

   \item Number of Daily Stressors (in the Current Week): It helps see how the most recent week went compared with the overall average.
\end{enumerate}

Collectively, Figure~\ref{fig:overall} provides an overview of overall trends and recent changes in their stress and stressors.

\subsection{Prominent Stressor Context}

To help identify their most dominant stressors, the second visualization in Week 1 (see Figure~\ref{fig:Prominent}) showed participants the locations and times of day associated with the five most prominent stressors. By tapping on each stressor, participants could view detailed information about the specific location and time of occurrence for the selected stressor.

\section{Week 2: Map View}

\begin{figure}[b]
     \centering
     \begin{subfigure}[b]{0.43\textwidth}
         \centering
         \includegraphics[width=\textwidth,height=6 cm]{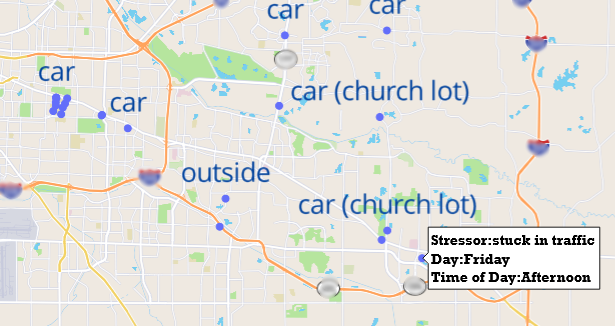}
\caption{Semantic locations of stressors on a map}
         \label{fig:map}
     \end{subfigure}
     \hfill
      \begin{subfigure}[b]{0.48\textwidth}
         \centering
         \includegraphics[width=\textwidth]{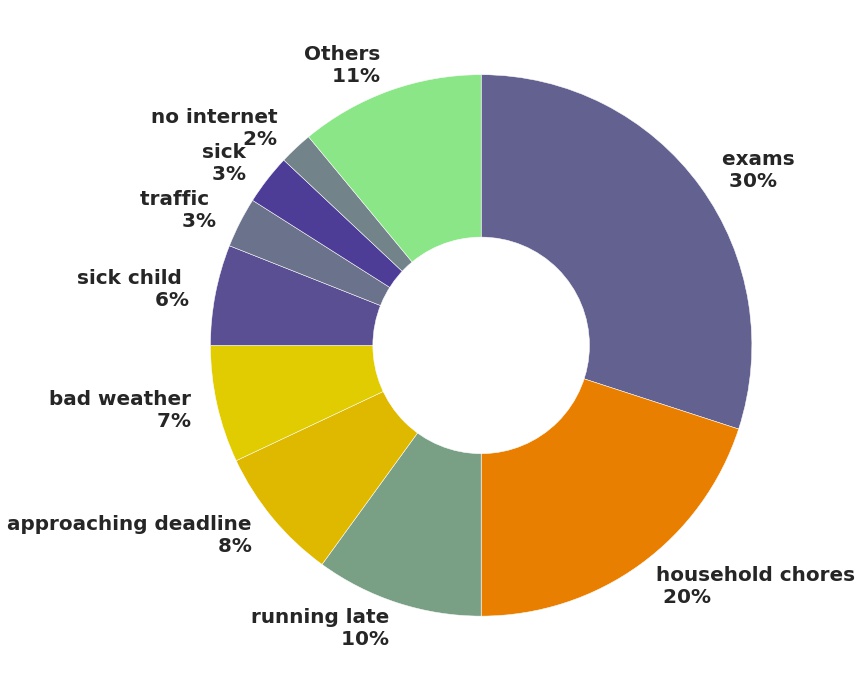}
\caption{Prevalence of prominent stressors }
         \label{fig:Stressor_prm}
     \end{subfigure}
     \hfill

        \caption{Visualizations for Weeks 2 and 3}
        \label{fig:vis_2_3}
        \Description{
}
\end{figure}

Using the GPS data and participants' reported semantic locations at the time of reporting a stressor, the map view (see Figure~\ref{fig:map}) showed semantic locations of reported stress instances on a map. Each point on the map corresponds to a specific location where stress was reported. By hovering over or tapping on each point, participants could access additional details such as the associated stressor, the day it occurred, and the specific time of day when the stressor was encountered. Furthermore, in cases where new locations were reported during the current week, participants had the option to interact with the map legend. Clicking or tapping on the legend allowed participants to access and explore the newly reported locations for the ongoing week, enhancing their understanding of stressor occurrences. 

\section{Week 3: Prevalence of Prominent Stressors}

Tens of stressors logged by the participants allowed an analysis of the prevalence of prominent stressors (i.e., the amount of time attributed to each stressor). Figure~\ref{fig:Stressor_prm} shows an example. Each segment of the chart represents a specific stressor, and participants had the option to interact with the visualization by hovering over or tapping on each segment. Upon interaction, the abbreviations representing the stressors expanded, revealing the corresponding percentage of stressed duration associated with that particular stressor. This interactive feature enabled the participants to gain insights into the relative impact of different stressors on their lives.

        \begin{figure}[t]
     \centering
     \begin{subfigure}[b]{0.43\textwidth}
         \centering
         \includegraphics[width=\textwidth]{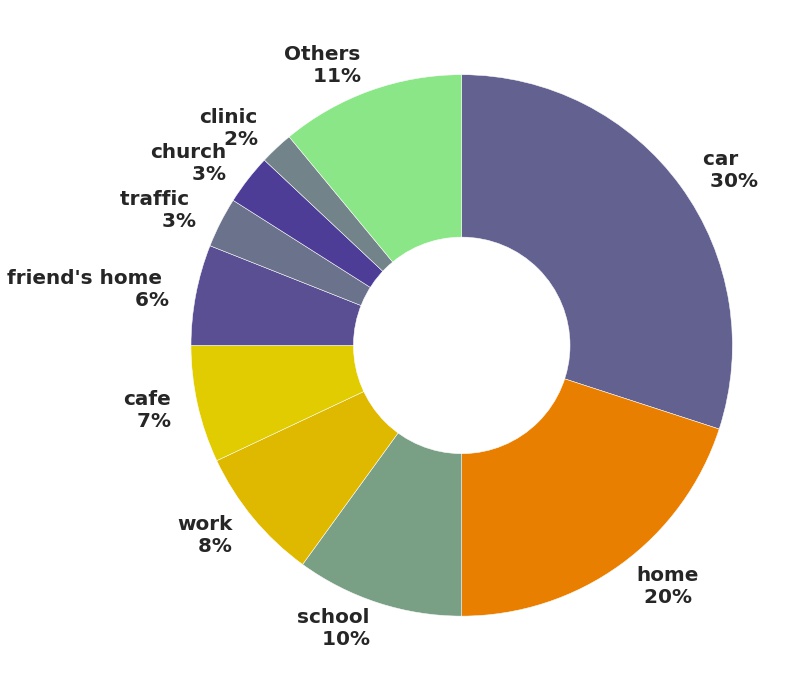}
\caption{Locations most associated with stress }
         \label{fig:loc_pro}
     \end{subfigure}
     \hfill
      \begin{subfigure}[b]{0.48\textwidth}
         \centering
         \includegraphics[width=\textwidth,height= 6 cm]{figs/calendar.png}
\caption{Calendar view of stressors}
         \label{fig:calendar}
     \end{subfigure}
     \hfill

        \caption{Visualizations for Weeks 4 and 5}
        \label{fig:week_5_6}
        \Description{
}
\end{figure}

\section{Week 4:  Locations Associated with Stress (Location Prominence)}
From the semantic location participants reported at the time of stress ratings and the duration of the event estimated by the app was used to determine the percent of total stressed time that participants were at specific locations. Figure~\ref{fig:loc_pro} an example of this visualization. Participants could engage with the visualization by clicking on or tapping the legend, which corresponded to specific locations. This action allowed participants to view the percentage of stressed duration associated with each location. By offering this interactive functionality, the visualization aimed to enhance participants' understanding of the spatial context of their stress experiences.

\section{Week 5: Calendar View of Stressors}
A calendar view can help participants associate stressors with their calendar and make it easier to identify the stress impact of various activities.
Therefore, we overlaid the reported stressors in a calendar view. Figure~\ref{fig:calendar} shows an example. It showed the days and times when specific stressors were reported and the color indicated the likelihood of the event determined to be stressful by the app. By hovering over or tapping on each segment within the visualization, participants could access additional details, including an expanded form of the abbreviation, the corresponding stress likelihood, and the average minutes reported for that hour. To facilitate interpretation, a color bar is situated on the right side of the visualization, indicating the stress likelihood values associated with different segments. 

\section{Week 6: Stressor Ranking}
\begin{figure}[t]
\includegraphics[width=\textwidth]{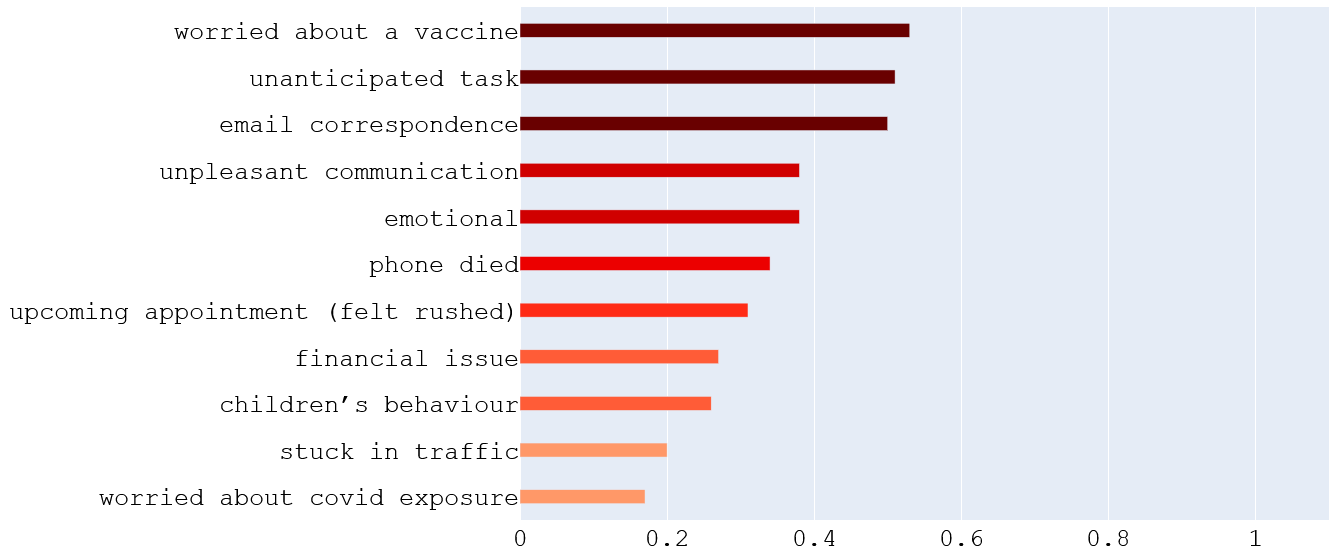}
\caption{Stressors associated with the highest stress likelihoods (Week 6) }
         \label{fig:ranking_stre}
\end{figure}

Each event presented to participants for stress rating and annotation has associated with it a likelihood with which the app determined this event to be indicative of stress. If the participant chose to denote a presented event as stressful and provided a stressor, this stressor becomes associated with the stress likelihood computed by the app for this event. Different stressors differ in the average stress likelihood associated with them. We used these to rank the stressors and showed the average stress likelihood associated with the highest-ranked stressors. Figure~\ref{fig:ranking_stre} shows an example of such a visualization. The stress likelihood displayed here was computed from the app and not from the stress ratings to allow for the possibility that in future apps, the burden of logging can be reduced further by having participants simply choose a stressor and not have to provide a stress rating each time they are prompted.

\begin{figure}[t]
\includegraphics[width=\textwidth]{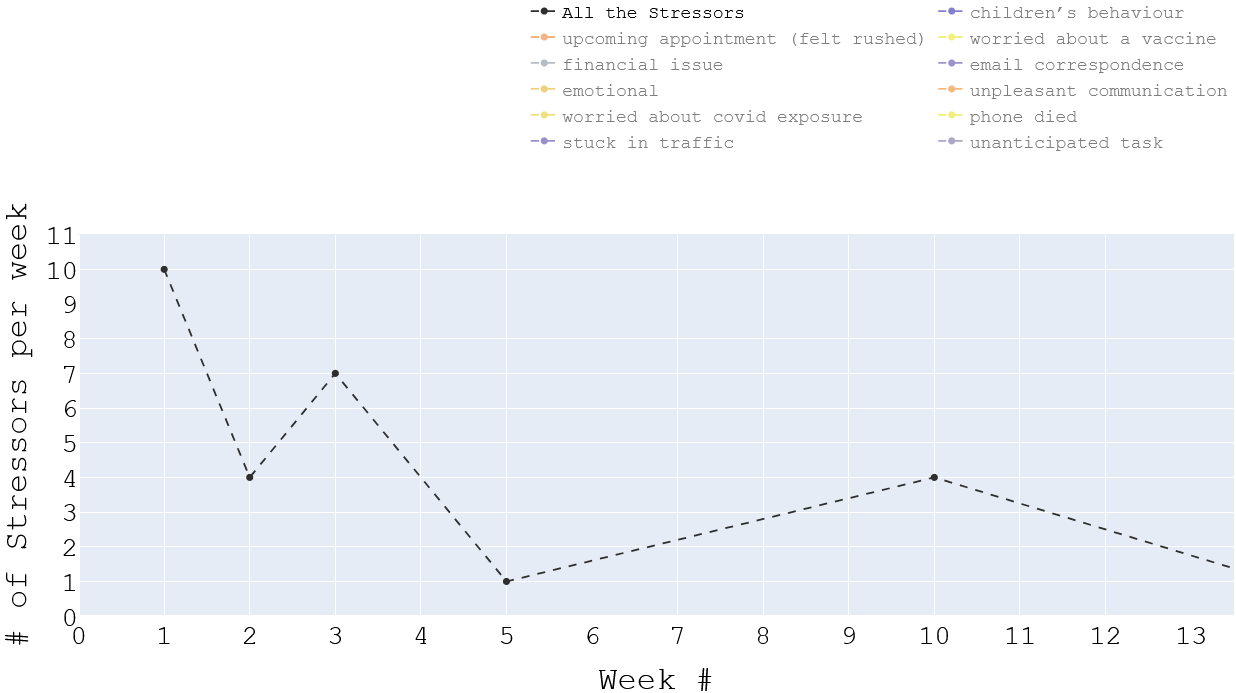}
\caption{Weekly trend in specific stressors reported per week over time (Week 7) }
         \label{fig:weekly_trend}
\end{figure}
\section{Week 7: Weekly Trend in Specific Stressors Reported Per Week}
Different weeks may differ in which stressors are reported and how frequently. The frequency of reports for some stressors may also change if the participants make some changes to their behaviors and are successful in mitigating them. To give them the ability to see such trends in their stressor logs, we provided them with a way to visually see the frequency of different stressors they reported. Figure~\ref{fig:weekly_trend} shows an example. Participants had the option to toggle stressors on and off in order to selectively view their representation on the graph. 

\section{Week 8: Weekly Prevalence}
\begin{figure}[t]
\includegraphics[width=0.9\textwidth,]{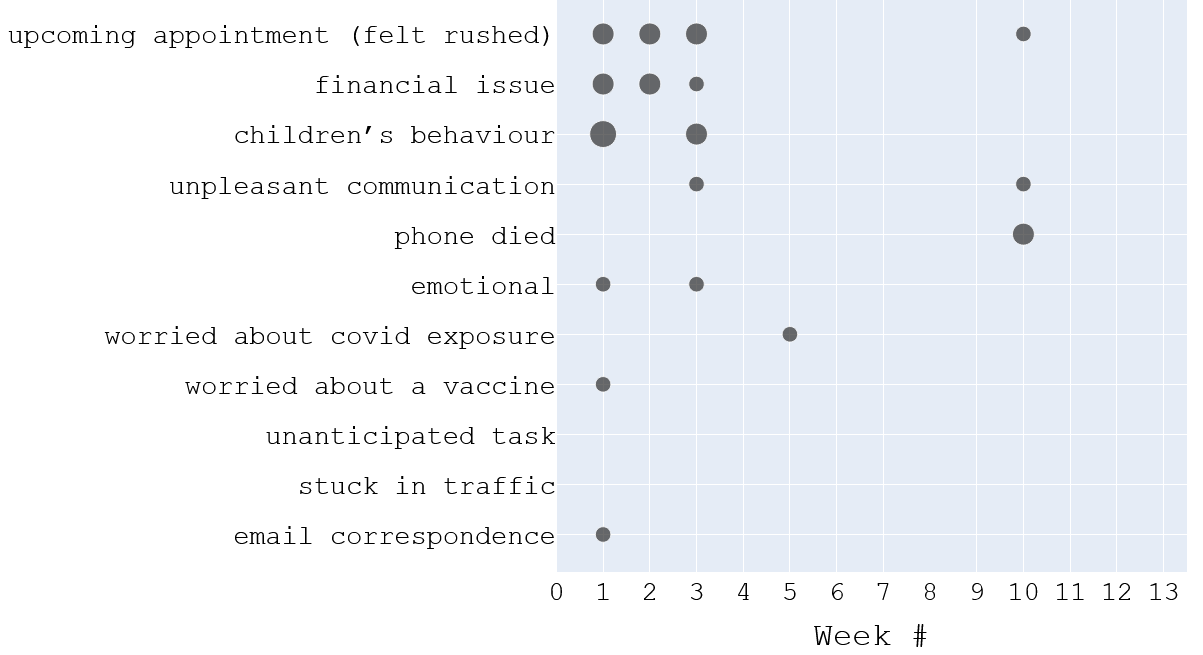}
\caption{Relative frequency with which specific stressors occur each week over time (Week 8) }
         \label{fig:weekly_prev}
\end{figure}

We then provided an alternative way to view the frequency of stressors reported each week. The goal was to allow them to compare the frequency of different stressors on the same graph so they could compare the relative frequency with which specific stressors were reported each. We wanted the graph to allow a comparison of frequency for different stressors for a given week and see how this changed over time. Figure~\ref{fig:weekly_prev} shows an example. The size of the dots represents the relative prevalence. Each segment on the graph corresponds to a specific stressor, and hovering or tapping over a segment provides additional information, including the expanded form of the abbreviation, the week number, and the corresponding number of occurrences for that stressor. 

\section{ Week 9: Time of day Trend }
\begin{figure}[t]
\includegraphics[width=0.9\textwidth]{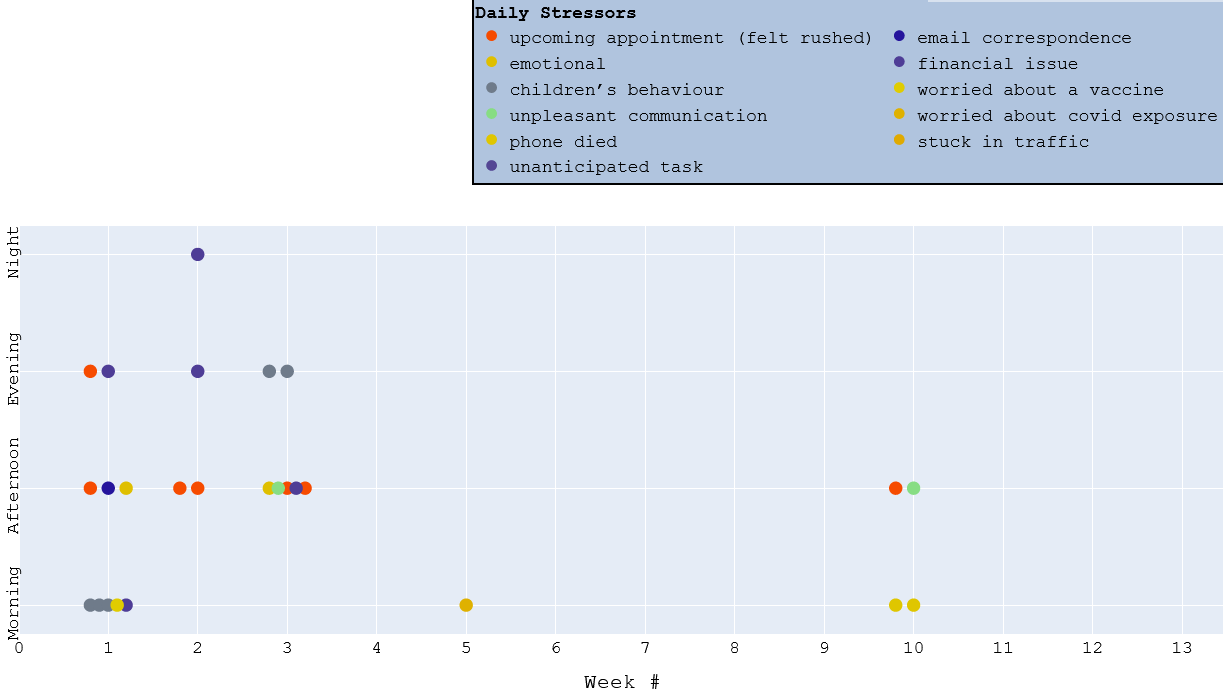}
\caption{Temporal patterns in stressor logs (Week 9) }
         \label{fig:time_trend}
\end{figure}

We wanted to provide a temporal context for different stressors so that participants could determine if there were specific temporal dynamics for some stressors. For this purpose, we provided them with a visualization to see the specific blocks of the day when different stressors occur. Figure~\ref{fig:time_trend} shows an example. It shows the blocks of time when specific stressors occur each week. The goal was to also allow them the ability to see any evolution in the temporal patterns over time. By interacting with the visualization, participants could select individual stressors by clicking or tapping on the corresponding legend item. This feature enabled participants to focus on specific stressors and gain insights into their temporal distribution throughout the MOODS study period. 

\section{Week 10: Location Trend}
\begin{figure}[h]
\includegraphics[width=0.9\textwidth]{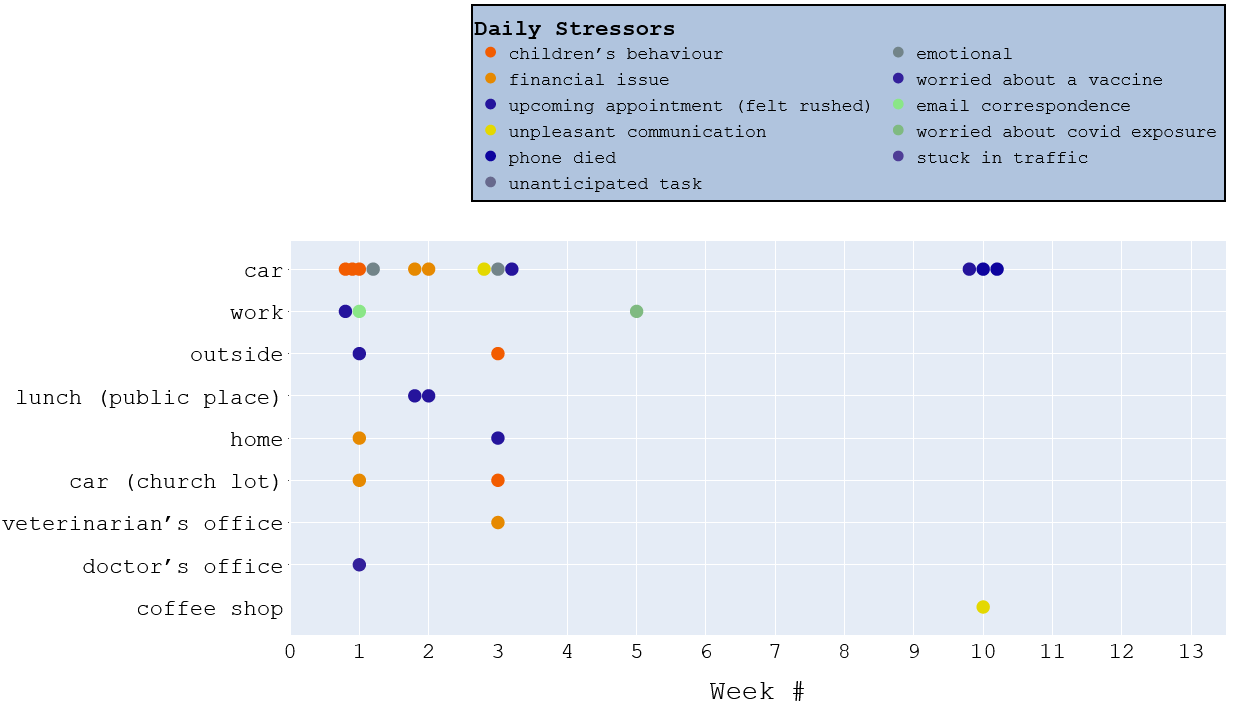}
\caption{Spatial patterns in stressor logs (Week 10)}
         \label{fig:loc_trend}
\end{figure}

Similar to facilitating the discovery of any temporal patterns in their data, we also wanted to facilitate the discovery of any spatial patterns in stressors. For this purpose, we revised the visualization from Week 9 by replacing the vertical axis to show the reported semantic locations instead of blocks of time. Figure~\ref{fig:loc_trend} shows an example of such a scatter plot. To focus on specific stressors, participants could interact with the graph by clicking or tapping the corresponding items in the legend. This interactive functionality allowed the participants to isolate and explore individual stressors, gaining insights into their occurrence patterns across various locations over the duration of the MOODS study.

\begin{figure}[h]
\includegraphics[width=\textwidth]{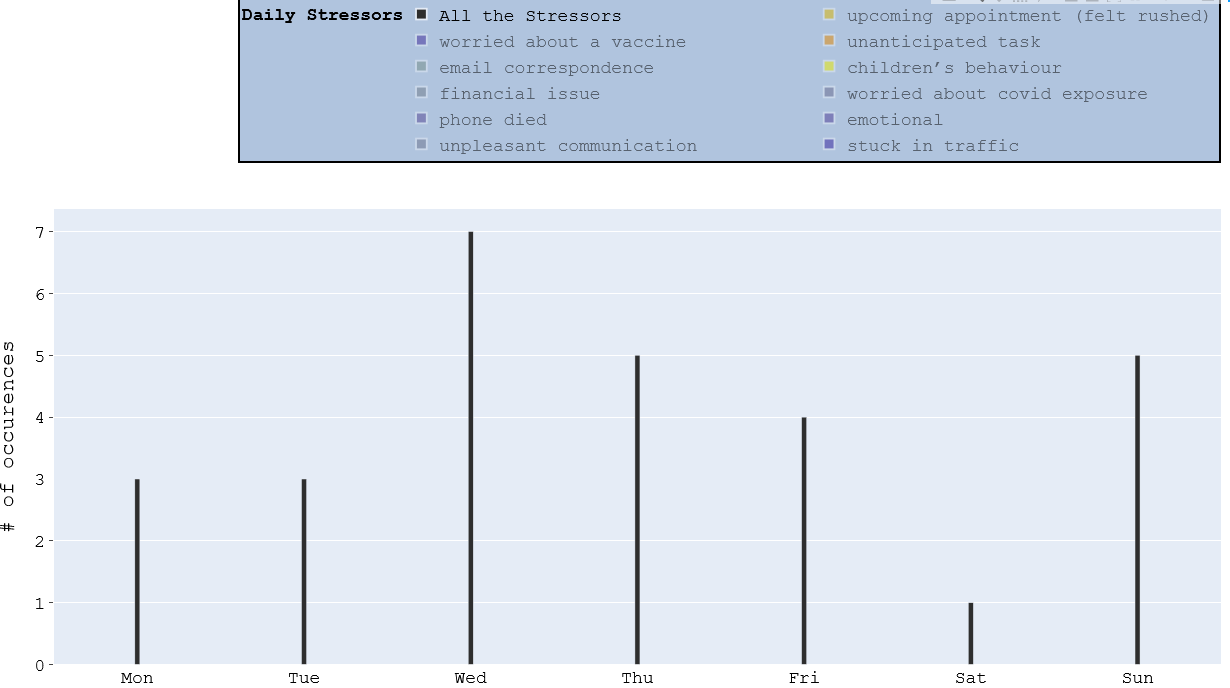}
\caption{Day of week patterns in stressor logs (Week 11)}
         \label{fig:day_trend}
\end{figure}

\section{Week 11: Day of the Week Patterns (Daily Trend)}
Some participants may have specific days of the week that may be more stressful than others depending on the nature of their work or personal life.  Figure~\ref{fig:day_trend} shows an example of a visualization that represents the frequency of reported stressors across different days of the week, spanning from the commencement of the study to the present date. In addition to showing which days of the week had the most stressors reported, it also allowed the ability to see trends in specific stressors. By interacting with the graph and clicking or tapping on the corresponding items in the legend, the participants could isolate individual stressors for closer examination. This interactive feature enabled participants to explore how specific stressors had been reported and distributed over various days of the week throughout the MOODS study.

\section{Week 12: Stressor Duration}
\begin{figure}[t]
\includegraphics[width=0.8\textwidth]{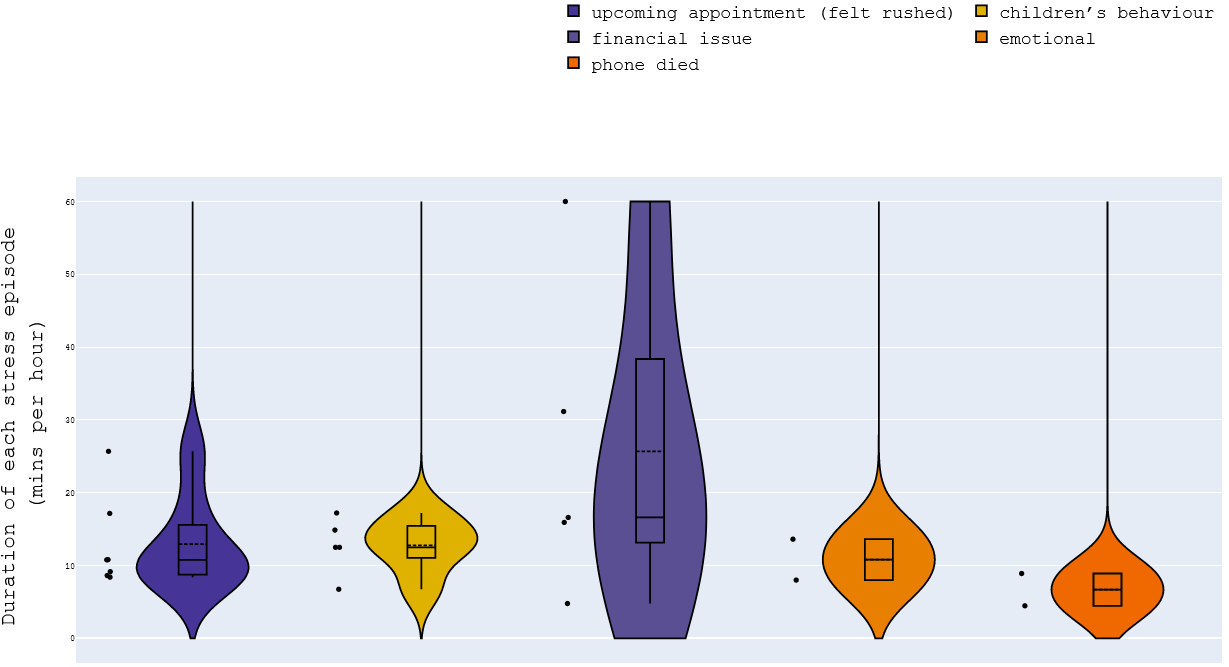}
\caption{Distribution of the duration of stress episodes associated with the five most prevalent stressors (Week 12) }
         \label{fig:vioil}
\end{figure}

The app estimated the duration of each bodily event that was presented to participants for stress rating and stressor annotation. Wa wanted to allow the participants to see if stressors differed in terms of the duration of each episode attributed to them. For this purpose, we presented a series of violin plots illustrating the duration distribution of the five most frequently reported stressors.  Figure~\ref{fig:vioil} shows an example. Each violin plot comprises box plots at their centers, providing insight into the central tendency and variability of the data. Additionally, the broader areas of the violins depict the probability density distribution across various values, which is smoothed using a kernel density estimator. 

\begin{figure}[b]
\includegraphics[width=\textwidth]{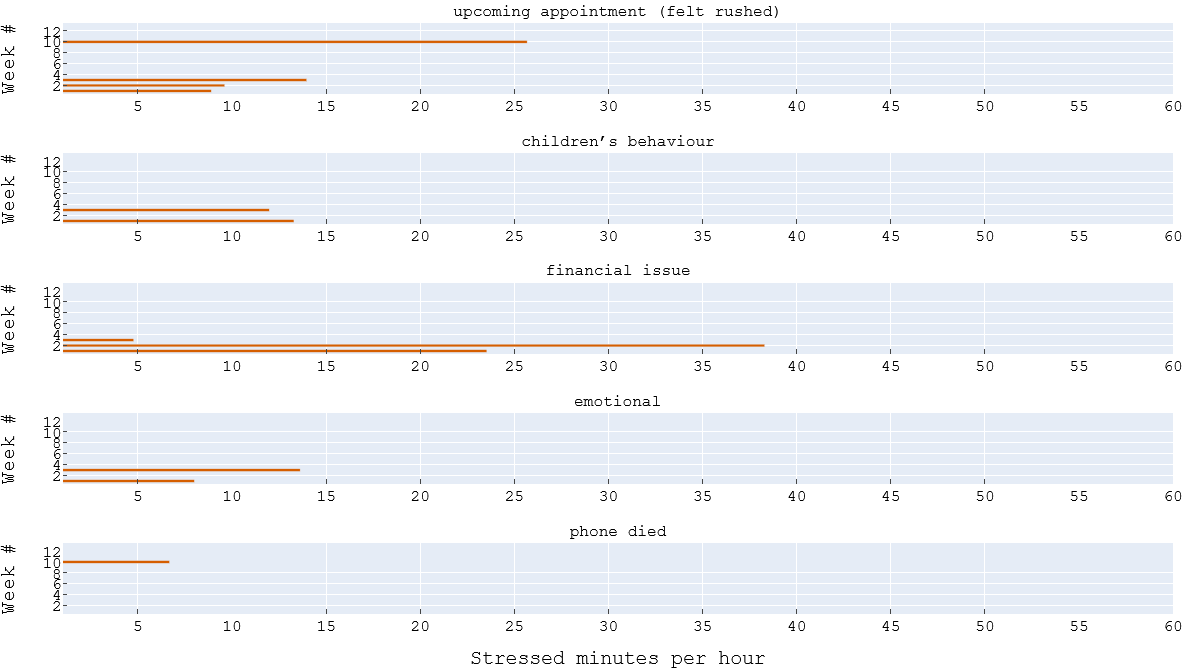}
\caption{Average stress episode duration associated with the five most prevalent stressors each week (Week 13) }
         \label{fig:duration}
\end{figure}

\section{Week 13: Prevalent Stressor Duration}
To dig deeper into the duration patterns, we next presented weekly patterns in the average duration per week for the top five reported stressors throughout the weeks of study participation. Figure~\ref{fig:duration} shows an example.  The visualization features bars, with red bars indicating the current week and gray bars representing previous weeks. This presentation allows participants to compare the duration of stressor periods over time, highlighting any variations or trends in stressed periods across different weeks.

\begin{figure}[h]
     \centering
     \begin{subfigure}[b]{0.45\textwidth}
         \centering
         \includegraphics[width=\textwidth]{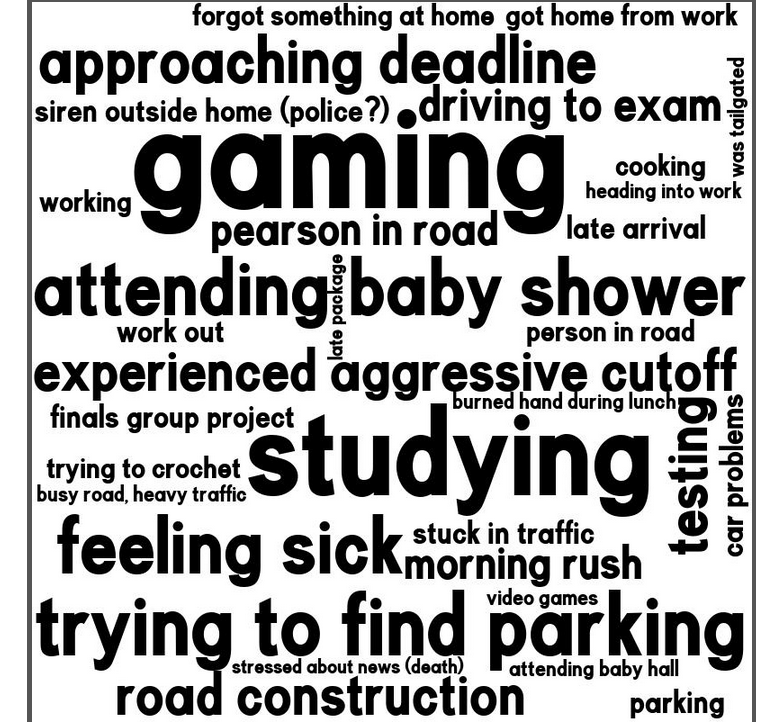}
         
         \captionsetup{justification=centering}
         \caption{Word clouds from the stressors log}         \label{fig:}
     \end{subfigure}
     \hfill
      \begin{subfigure}[b]{0.45\textwidth}
         \centering
         \includegraphics[width=\textwidth]{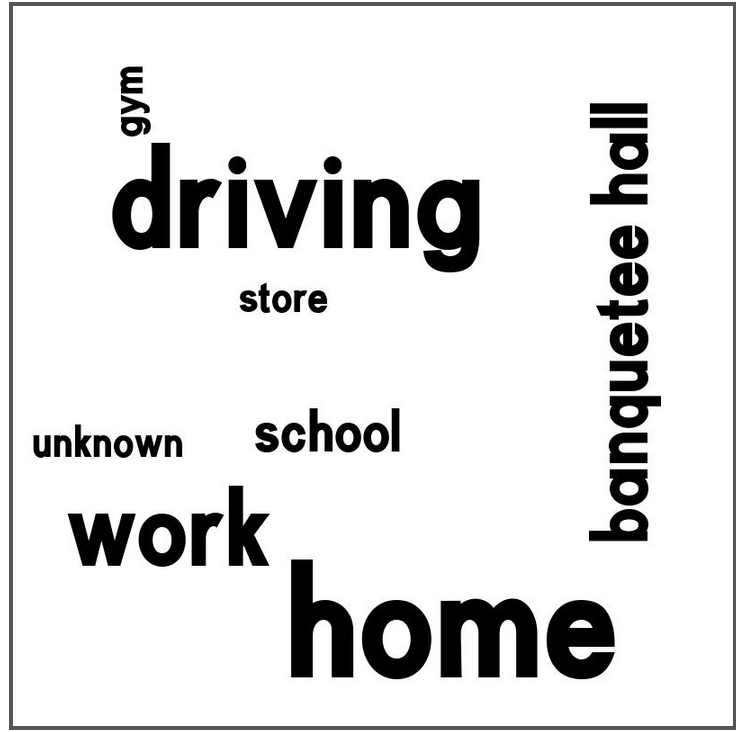}
         \captionsetup{justification=centering}
         \caption{Word clouds of locations reported with stressors}
         \label{fig:}
     \end{subfigure}
     \hfill

        \caption{Visualizations for Week 14}
        \label{fig:wc}
        \Description{
}
\end{figure}
\section{Week 14: Word Clouds}
 Finally, when we have the most amount of data in the stressor logs collected, we created Figure~\ref{fig:wc} illustrates word clouds to show patterns in the stressors reported and their associated reported locations. In this visualization, different stressors and locations are represented by words, with font sizes reflecting their frequency of occurrence. Larger fonts indicate more frequently reported stressors or locations. This word cloud offers a quick visual insight into the relative prevalence of different stressors and the locations associated with reported stressors.
